\documentclass[aps,onecolumn,preprint,superscriptaddress,nofootinbib,floats,tightenlines,12pt]{revtex4}
\pdfoutput=1

\usepackage{amsmath,amssymb,color,mathrsfs,verbatim,epsfig, bbm, wasysym}
\usepackage[hyperfootnotes=false]{hyperref}
\usepackage{slashed}
\usepackage{graphicx}

\allowdisplaybreaks

\begin{document}

\baselineskip=18pt


\thispagestyle{empty}
\vspace{20pt}
\font\cmss=cmss10 \font\cmsss=cmss10 at 7pt

\begin{flushright}
\small 
\end{flushright}

\hfill
\vspace{20pt}

\begin{center}
{\Large \textbf
{
Group Theoretic Approach
\\[0.15cm] 
to Fermion Production
}}
\end{center}

\vspace{15pt}
\begin{center}
{Ui Min$^{\, a}$, Minho Son$^{\, a}$ and Han Gyeol Suh$^{\, a}$}
\vspace{30pt}

$^{a}$ {\small \it Department of Physics, Korea Advanced Institute of Science and Technology, \\ 291 Daehak-ro, Yuseong-gu, Daejeon 34141, Republic of Korea
}

\end{center}

\vspace{20pt}
\begin{center}
\textbf{Abstract}
\end{center}
\vspace{5pt} {\small
We propose a universal group theoretic description of the fermion production through any type of interaction to scalar or pseudo-scalar.  Our group theoretic approach relies on the group $SU(2) \times U(1)$, corresponding to the freedom in choosing representations of the gamma matrices in Clifford algebra, under which a part of the Dirac spinor function transforms like a fundamental representation. In terms of a new $SO(3)$ ($\sim SU(2)$) vector constructed out of spinor functions, we show that fermion production mechanism can be analogous to the classical dynamics of a vector precessing with the angular velocity. In our group theoretic approach, the equation of motion takes a universal form for any system, and choosing a different type of interaction or a different basis amounts to selecting the corresponding angular velocity. The expression of the particle number density is greatly simplified, compared to the traditional approach, and it provides us with a simple geometric interpretation of the fermion production dynamics. For the purpose of the demonstration, we focus on the fermion production through the derivative coupling to the pseudo-scalar.}

\vfill\eject
\noindent

\tableofcontents
\newpage

\section{Introduction}
\label{sec:intro}
The particle production is an efficient way of dissipating energy, and it has a variety of application from the phenomenology to the cosmology.  In cosmology, the particle production has been known to be an underlying mechanism,  known as the preheating via parametric resonance or excitation, that is responsible for the reheating of the Universe in the post-inflationary era~\cite{Kofman:1997yn}. The axion inflation through the particle production has been explored. For instance, the axion through the electromagnetic dissipation can be realized as the inflaton even in the steep axion potential~\cite{Anber:2009ua}. The fermion production could be significant enough, or more efficient than the dissipation via the Hubble friction (against a common prejudice) to support the axion inflation as well~\cite{Adshead:2015kza,Adshead:2018oaa}. The particle production could also generate the gravitational waves in various context~\cite{Khlebnikov:1997di,Easther:2006gt,Easther:2006vd,Easther:2007vj,GarciaBellido:2007dg,Dufaux:2007pt,Dufaux:2010cf,Bethke:2013aba,Figueroa:2013vif,Bethke:2013vca,Figueroa:2016ojl,Figueroa:2017vfa,Adshead:2018doq}.
In phenomenology, an interesting recent application is the relaxation with the particle production~\cite{Hook:2016mqo}~\footnote{See~\cite{Fonseca:2018xzp} for the discussion about the Higgs production as an alternative to the gauge boson production for the dissipation, and also for the phenomenological and cosmological constraints on the relaxation model proposed in~\cite{Hook:2016mqo}.} that has been proposed as an alternative solution to the gauge hierarchy problem~\cite{Graham:2015cka}.

In this work, we revisit the theory of the spin-1/2 fermion production~\cite{Greene:1998nh,Greene:2000ew,Peloso:2000hy,GarciaBellido:2000dc,Tsujikawa:2000ik,Garbrecht:2002pd,Adshead:2015kza,Adshead:2018oaa,Domcke:2018eki} and reformulate it in a group theoretic way.  Our formalism is based on the reparametrization group that corresponds to the freedom in choosing a representation of the gamma matrices. Since this freedom is  unphysical, the physical observables must be invariant under the reparametrization group. In a typical quantum field theory, one chooses a representation of the gamma matrices in the beginning as a convention, and therefore, the corresponding freedom is hidden and it can hardly become practical. As we will demonstrate in this work, we newly discover that the freedom in the representation of the gamma matrices can greatly help us understanding the complicated fermion production mechanism. As an example, our group theoretic approach reveals a simple analogy between the quantum-mechanical dynamics of the fermion production and the classical dynamics of a vector precessing with an angular velocity. This analogy provides us with a simple geometrical interpretation of the quantum-mechanical fermion production. To the best of our knowledge, we are the first who show that the reparametrization group in the representation of the gamma matrices could be useful for understanding the nature of the fermion production.

This paper is organized as follows. In Section~\ref{sec:model}, we set up the model for the fermion production through the coupling of the pseudo-scalar to fermions, and we discuss about subtleties caused by the basis choice. 
In Section~\ref{sec:grouptheoryapproach}, we first establish the existence of the reparametrization group that leaves Clifford algebra and the Lagrangian for the Dirac fermion invariant and that does not overlap with the Lorentz group. Then, we construct the building blocks for our group theoretic approach such as the irreducible representations of the reparametrization group. In Section~\ref{sec:Fermion:Inertialframe}, we reformulate the equations of motion of fermions and particle number density in an inertial frame in terms of the covariant or invariant quantities under the reparametrization group. We demonstrate the analogy of the fermion production mechanism to the classical motion of a vector precessing with an angular velocity. In Section~\ref{sec:Fermion:rotatingframe},  we demonstrate how the formalism changes under the time-dependent transformation from an inertial frame to the non-inertial frame.
In Section~\ref{sec:numeric:simulation}, we perform some numerical study to elaborate our new approach compared to the traditional way. In Section~\ref{sec:summary}, we summarize our results. In Appendix~\ref{sec:app:convention}, we provide the convention of the metric and the gamma matrices. In Appendix~\ref{sec:app:particlenumber}, we provide the explicit derivation of the particle number from the Hamiltonian in an inertial frame.

\section{The model}
\label{sec:model}

We study the fermion production through the derivative coupling of the Dirac fermion $\psi$ to a pseudo-scalar $\phi$ with the action,
\begin{equation}
 \mathcal {S} = \int d^4 x \sqrt{-g} \Big [ \bar{\psi}\left ( i e^\mu_{\ a} \gamma^a D_\mu - m - \frac{1}{f} e^\mu_{\ a} \gamma^a\gamma^5\partial_\mu \phi \right ) \psi + \frac{1}{2} (\partial_\mu \phi )^2 - V(\phi) \Big ]~,
\end{equation}
on a metric
\begin{equation}
  ds^2 = dt^2 - a(t)^2 d{\bf x}^2 = a(t)^2 \left ( d\tau^2 - d{\bf x}^2 \right )~,
\end{equation}
where $a(t)$ is a scale factor of the Universe. The overall scale factor due to $\sqrt{-g}$ in the Lagrangian for fermions can be removed via rescaling, $\psi \rightarrow a^{-3/2} \psi$. Under this rescaling, the covariant derivative due to the spin connection become partial derivative. The resulting Lagrangian becomes
\begin{equation}\label{eq:Lag:Y}
 \mathcal {L} = \bar{\psi}\left ( i \gamma^\mu \partial_\mu - m a - \frac{1}{f} \gamma^\mu\gamma^5\partial_\mu \phi \right ) \psi + \frac{1}{2} a^2 \eta^{\mu\nu}\partial_\mu \phi \partial_\nu \phi - a^4 V(\phi)~.
\end{equation}
Throughout this work we will use the symbol $\tau$ to denote the conformal time, and we will not distinguish the cosmic time and the conformal time unless necessary. We will also assume that the pseudo-scalar field $\phi$ is spatially homogeneous.

In the context of the fermion production during axion inflation, the analytic solution of the differential equation from the Lagrangian in Eq.~(\ref{eq:Lag:Y}) is available, known as the Whittaker function, assuming that the homogeneous $\phi$ has nearly constant velocity~\cite{Adshead:2015kza}. Even when assuming the static universe scenario, the apparent formalism for the fermion production has a close similarity to the case with the Yukawa-type coupling of the scalar to the fermions~\cite{Adshead:2015kza}, which could be useful for a better understanding. However, the corresponding Hamiltonian formalism is not straightforward to use to define the particle number density unambiguously~\cite{Adshead:2018oaa}. It is because the derivative coupling of the pseudo-scalar to fermions includes the velocity of the pseudo-scalar, $\dot\phi$, and this causes an extra fermion-bilinear term in its conjugate momentum:
\begin{equation}
  \Pi_\psi = \frac{\delta \mathcal{L}}{\delta \dot{\psi}} = i \psi^\dagger~, \quad
  \Pi_\phi = \frac{\delta \mathcal{L}}{\delta \dot{\phi}} = a^2 \dot{\phi} - \frac{1}{f}\bar{\psi}\gamma^0\gamma^5 \psi~.
\end{equation}
The Hamiltonian is obtained by the Legendre transformation,
\begin{equation}\label{eq:Y:hamiltonian}
\begin{split}
 \mathcal{H} &= \Pi_\psi\, \dot{\psi} + \Pi_\phi\, \dot{\phi} - \mathcal{L} \\
  &= \bar{\psi} \left ( - i \gamma^i \partial_i + m a+ \frac{1}{f} \gamma^0\gamma^5\dot{\phi} \right ) \psi 
  -\frac{1}{2 a^2} \frac{ \left ( \bar\psi \gamma^0 \gamma^5 \psi \right )^2}{f^2}
  +\frac{1}{2 a^2} \Pi_\phi^2
  + a^4 V(\phi)~,
\end{split}
\end{equation}
where we organized the Hamiltonian such that the first quadratic term in $\psi$ matches to the part taken as the free Hamiltonian in literature~\cite{Adshead:2015kza,Adshead:2018oaa}, from which the particle number was estimated. One notices that the remaining part of the Hamiltonian includes the four-fermion self interaction when expressed in terms of the conjugate momentum $\Pi_\phi$, and the zero particle production in the massless limit is not straightforward.

The estimation of the fermion production is more straightforward in the Hamiltonian formalism from the Lagrangian obtained via the field redefinition~\cite{Adshead:2018oaa}, 
\begin{equation}\label{eq:rot:Ytopsi}
  \psi \rightarrow e^{-i \gamma^5 \phi/f} \psi~.
\end{equation}
After the rotation in Eq.~(\ref{eq:rot:Ytopsi}), the Lagrangian becomes
\begin{equation}\label{eq:Lag:psi}
 \mathcal {L} = \bar{\psi}\left ( i \gamma^\mu \partial_\mu - m_R + i\, m_I \, \gamma^5 \right ) \psi + \frac{1}{2} a^2 \eta^{\mu\nu}\partial_\mu \phi \partial_\nu \phi - a^4 V(\phi)~.
\end{equation}
where $m_R = ma\, {\rm cos} \left ( \frac{2\phi}{f} \right ) $ and $m_I = ma\, {\rm sin} \left ( \frac{2\phi}{f} \right )$.  The conjugate momenta are derived to be
\begin{equation}\label{eq:conj:mom:psi}
  \Pi_\psi =  i \psi^\dagger~, \quad
  \Pi_\phi = a^2 \dot{\phi}~,
\end{equation}
and the Hamiltonian is given by
\begin{equation}\label{eq:psi:hamiltonian}
\begin{split}
 \mathcal{H} 
  &= \bar{\psi} \left ( - i \gamma^i \partial_i + m_R - i\, m_I\, \gamma^5 \right ) \psi +  \frac{1}{2} a^2 \dot{\phi}^2 + a^4 V(\phi)~.
\end{split}
\end{equation}
As a result, the fermion and pseudo-scalar parts in the Hamiltonian in Eq.~(\ref{eq:psi:hamiltonian}) are clearly separated, and the fermion Hamiltonian includes only the quadratic terms in $\psi$. In the Hamiltonian in Eq.~(\ref{eq:psi:hamiltonian}), the decoupling of the pseudo-scalar from the fermions in the massless limit is manifest. The fermion becomes a free field in the massless limit, and therefore, no fermion is produced.

In our approach, we will stick to the basis in which the Lagrangian and Hamiltonian take the forms without the derivative couplings (see Eqs.~(\ref{eq:Lag:psi}) and~(\ref{eq:psi:hamiltonian})) and develop our group theoretic approach. After we construct our approach in one basis, we will discuss about how the fermion production dynamics changes when switching from one basis to another basis with the derivative coupling.

\section{Reparametrization Group}
\label{sec:grouptheoryapproach}

To establish the reparametrization group later, with a clear comparison with the Lorentz group, that our group theoretic approach is based on, we will start with briefly reviewing the spinor representation of the Lorentz group. Our starting point is the Clifford algebra,
\begin{equation}\label{eq:clifford}
\{ \gamma^\mu ,\, \gamma^\nu \} = 2\eta^{\mu\nu}\, I_{\bf 4}~,
\end{equation}
where $I_{\bf n}$ denotes $n\times n$ identity matrix. The gamma matrices in the Weyl representation are suitable for the discussion of the Lorentz group, and they are given by (also in the tensor product form of two $2\times 2$ matrices)
\begin{equation}
\gamma^0 
= \begin{pmatrix}  0 & I_{\bf 2} \\ I_{\bf 2} & 0 \end{pmatrix} 
= \sigma_1 \otimes I_{\bf 2}~,~
\gamma^i  
= \begin{pmatrix}  0 &  \sigma_i \\ -\sigma_i & 0 \end{pmatrix} 
= i\, \sigma_2 \otimes \sigma_i~,~
\gamma^5 
= \begin{pmatrix}  -I_{\bf 2} & 0 \\ 0 & I_{\bf 2} \end{pmatrix} 
= - \sigma_3 \otimes I_{\bf 2}~,
\end{equation}
where $\otimes$ refers to the tensor product whereas $\oplus$ is used to refer to the tensor sum.
The spinor representation of the Lorentz group is defined as the following commutator of two gamma matrices in the Clifford algebra,
\begin{equation}
S^{\mu\nu} = \frac{i}{4}[\gamma^\mu,\, \gamma^\nu]~,
\end{equation}
and it satisfies the Lorentz algebra.  The six generators of $S^{\mu\nu}$ can be split into three space rotations and three Lorentz boosts:
\begin{equation}\label{eq:spinorrep:spacerot}
J_i \equiv \frac{1}{2} \epsilon_{ijk}S^{jk} = \frac{1}{2} I_{\bf 2} \otimes \sigma_i~,\quad
K_i \equiv  S^{i0} = \frac{i}{2} \sigma_3 \otimes \sigma_i~.
\end{equation} 
The generators in Eq.~(\ref{eq:spinorrep:spacerot}) can be reorganized to satisfy two independent $SU(2)$ Lie algebras:
\begin{equation}\label{eq:JL/R}
\left(J_{L/R}\right)_i \equiv \frac{J_i \mp i K_i}{2} = \frac{1}{2}\left ( I_{\bf 2} \pm \sigma_3 \right ) \otimes \frac{\sigma_i}{2}~.
\end{equation}
One sees that the Lorentz group is isomorphic to $SU(2)_L \times SU(2)_R$ whose Casimir operators are used to construct the irreducible representations of the Lorentz group. The four-component Dirac spinor belongs to the $(1/2,\, 0)\oplus (0,\, 1/2)$ representation of $SU(2)_L \times SU(2)_R$, and it can be written as
\begin{equation}\label{eq:spinor:LRform}
\psi = \begin{pmatrix} \psi_L \\[3pt] \psi_R  \end{pmatrix}~.
\end{equation}
On the other hand, the action of the group element for the space rotation with the generators in Eq.~(\ref{eq:spinorrep:spacerot}) is manifest in the following tensor product form,
\begin{equation}\label{eq:spinor:tensorform}
\psi = \xi \otimes \chi~,
\end{equation} 
where $\xi$ denotes two-component column vector and $\chi$ two-component spinor. The tensor product form in Eq.~(\ref{eq:spinor:tensorform}) should be understood to hold for a Fourier mode of the Dirac spinor as its meaning will be clear below. The space rotation acts on the Dirac spinor $\psi$ like
\begin{equation}\label{eq:spacerot:on:spinor}
\psi \rightarrow e^{-i \vec{\theta}\cdot \vec{J}}\, \psi =  \xi \otimes e^{-i \vec{\theta}\cdot \frac{\vec{\sigma}}{2}}\, \chi~,
\end{equation}
which implies that the space rotation rotates $\psi_L$ and $\psi_R$ universally whereas the Lorentz boosts are not associated with any rotation. The space rotation in Eq.~(\ref{eq:spacerot:on:spinor}) will be compared with the subgroup of the reparametrization group below.

The representation of the gamma matrices is not unique. Indeed, the Clifford algebra in Eq.~(\ref{eq:clifford})
is invariant under a similarity transformation,
\begin{equation}\label{eq:simtr:SL4C}
\gamma^\mu \rightarrow U \gamma^\mu U^{-1}~,
\end{equation}
with an $4\times 4$ unitary matrix $U$. Although the maximal transformation group that keeps algebra invariant is the complex general linear group $GL(4,{\bf C})$, unitarity condition is required to keep Dirac theory invariant at the same time. 

We consider the following subgroup of $U(4)$, which was constructed by tensor products of two unitary matrices and phase rotation,
\begin{equation}
SU(2)_1 \times SU(2)_2 \times U(1)\subset U(4)~.
\end{equation}
The $U(1)$ is the global phase transformation. In the parametrization of the Dirac spinor like Eq.~(\ref{eq:spinor:tensorform}), we will assume that $\xi$ carries $U(1)$ charge. The matrix representation of $SU(2)_1\times SU(2)_2$ is obatined by tensor product, and it acts on the Dirac spinor in Eq.~(\ref{eq:spinor:tensorform}) like
\begin{equation}\label{eq:productgroup:on:spinor}
\psi = \xi \otimes \chi \rightarrow \left (U_1 \otimes U_2 \right )  \left ( \xi \otimes \chi \right ) = (U_1\xi) \otimes (U_2 \chi)~
\end{equation}
where $U_1$ ($U_2$) is the matrix representation of $SU(2)_1$ ($SU(2)_2$). The $U_2$ transformation universally acts on two-component spinors in $\psi_{L}$ and $\psi_{R}$.  Although the space rotation of the Lorentz group in Eq.~(\ref{eq:spacerot:on:spinor}) and the $U_2$ transformation on the spinor in Eq.~(\ref{eq:productgroup:on:spinor}) look similar, the $SU(2)_2$ group can not be identified with $SU(2)$ for the space rotation of the Lorentz group~\footnote{The group properties on the gamma matrices and the transformation acting on the Dirac spinor are identical for both $SU(2)_2$ and $SU(2)$ space rotation of the Lorentz group. However, the gamma matrices do not transform under the Lorentz transformation, $\gamma^\mu \rightarrow \gamma^\mu$, whereas the gamma matrices transform under the similarity transformation of the reparametrization group as is indicated in Eq.~(\ref{eq:simtr:SL4C}). As a result, for instance, a vector current $\bar{\psi}\gamma^\mu \psi$ stays invariant under the reparametrization transformation whereas it transforms like a vector under the Lorentz group.}. We do not find any relevant role played by the $SU(2)_2$ subgroup in our work, and therefore, we will not consider it anymore.



The $U_1$ transformation of $SU(2)_1$ exchanges between $\psi_{L}$ and $\psi_{R}$, and it does not overlap with the Lorentz group.  A well-known example of $SU(2)_1$ is the similarity transformation between Weyl and Dirac representations of the gamma matrices. The rotation by $\pi/2$  about $x_2$-axis of $SU(2)_1$ transforms the gamma matrices in Weyl representation into those in the Dirac representation:
\begin{equation}
\gamma^0 
= \begin{pmatrix} I_{\bf 2} & 0 \\ 0 & - I_{\bf 2} \end{pmatrix} 
= \sigma_3 \otimes I_{\bf 2}~,~
\gamma^i  
= \begin{pmatrix}  0 &  \sigma_i \\ -\sigma_i & 0 \end{pmatrix} 
= i\, \sigma_2 \otimes \sigma_i~,~
\gamma^5 
= \begin{pmatrix}  0 & I_{\bf 2} \\ I_{\bf 2} & 0 \end{pmatrix} 
= \sigma_1 \otimes I_{\bf 2}~.
\end{equation}
Since $SU(2)_1 \times U(1)$ is a symmetry in choosing the representation of the gamma matrices, any physical quantity should be invariant under the symmetry. Importantly, the two-component column vector $\xi$ in Eq.~(\ref{eq:spinor:tensorform}) transforms like the fundamental representation of $SU(2)_1$ with a charge under $U(1)$. Our group theoretic construction of the fermion production relies on this property.

For the discussion of the fermion production through the coupling to the pseudo-scalar, we need to quantize the Dirac spinor in the Lagrangian while keeping pseudo-scalar as a classical field. A generic fermion quantum field can be written as
\begin{equation}\label{eq:psi:quantum}
\psi = \int \frac{d^3k}{(2\pi)^{3/2}} e^{i{\bf{k}}\cdot {\bf{x}}} \sum_{r=\pm} \left [ U_r( {\bf{k}}, \tau) a_r ({\bf{k}}) + V_r(-{\bf{k}},\tau) b^\dagger_r (-{\bf{k}})  \right ]~.
\end{equation}
A Fourier mode in Eq.~(\ref{eq:psi:quantum}) is what we actually meant in Eq.~\eqref{eq:spinor:tensorform}. The spinor function $U_r$ can be written in the tensor product form,
\begin{equation}
U_r({\bf{k}},\tau) = \frac{1}{\sqrt{2}} \begin{pmatrix} u_r \chi_r \\ v_r\, r\, \chi_r \end{pmatrix}
= \frac{1}{\sqrt{2}} \begin{pmatrix} u_r \\ r\, v_r \end{pmatrix} \otimes \chi_r 
\equiv \xi_r({\bf{k}} ,\tau) \otimes \chi_r({\bf{k}})~,
\end{equation}
where $\xi_r$ is a $SU(2)_1$ doublet which carries the $U(1)$ charge and $\chi_r$ is a helicity eigenstates corresponding to the momentum vector $\bf{k}$. The other spinor function $V_r$ is related to $U_r$ via the charge conjugation (see Appendix~\ref{sec:app:particlenumber} for the detail). 

As a first step to construct physical parameters, we construct the bilinear of $\xi$. Due to the $U(1)$ invariance, it takes the form,
\begin{equation}
\xi_r^\dagger\, A\, \xi_r~
\end{equation}
where $A$ is an arbitrary $2\times 2$ complex matrix. Since an arbitrary $2\times 2$ complex matrix can be written as a linear combination of $I_2$ and $\sigma_i$, the only $U(1)$ invariant $\xi$ bilinears are, in terms of $SO(3)_1 \sim SU(2)_1$,
\begin{equation}\label{eq:SO3:tensors}
\begin{array}{cll}
\xi_r^\dagger \xi_r & : & \hbox{scalar}~, 
\\[5pt]
\xi_r^\dagger \sigma_i\, \xi_r & : & \hbox{vector}~.
\end{array}
\end{equation}
The $SO(3)_1$ scalar is just a normalization. We normalize it to $\xi_r^\dagger \xi_r = (|u_r|^2 + |v_r|^2)/2 = 1$. The only non-trivial representation for the spin-1/2 fermion production is the $SO(3)_1$ vector, and we define it as $\vec{\zeta}_r$:
\begin{equation}
\vec{\zeta}_r \equiv \xi^\dagger \sigma_i\, \xi  = \displaystyle\frac{1}{2} \left ( u^*_r, \, r\, v^*_r \right ) \vec{\sigma}  \begin{pmatrix} u_r  \\[2pt]  r\, v_r \end{pmatrix}~.
\end{equation}
The explicit form of three components of $\vec{\zeta}_r$ is
\begin{equation}\label{eq:zeta:comp}
\begin{split}
\zeta_{r\, 1} = \frac{1}{2}r \left ( u^*_r v_r + v^*_r u_r \right ) ~,\quad
\zeta_{r\, 2} = -\frac{1}{2} i\, r \left ( u^*_r v_r - v^*_r u_r \right )~,\quad
\zeta_{r\, 3} = \frac{1}{2}\left ( |u_r|^2 - |v_r|^2 \right ) ~,
\end{split}
\end{equation}
which appeared in many places in~\cite{Adshead:2018oaa}. Note that $\vec{\zeta}_r$ is real unit vector, or $|\vec{\zeta}_r|=1$, with our normalization of $\xi^\dagger \xi = 1$.

\section{Fermion Production}
\subsection{Fermion Production in Inertial Frame}
\label{sec:Fermion:Inertialframe}
As was mentioned in Section~\ref{sec:model}, defining the particle number is more straightforward with the Lagrangian in Eq.~(\ref{eq:Lag:psi}). We will call this basis an inertial frame, borrowing the terminology from the classical mechanics, to distinguish it from another $\vec{\zeta}_r$ frame that will be introduced in Section~\ref{sec:Fermion:rotatingframe}. Its meaning will be clear as we will develop the analogy of the fermion production to the system in classical mechanics.

The equation of motion for $\psi$ from the Lagrangian in Eq.~(\ref{eq:Lag:psi}) is given by
\begin{equation}
  \left ( i\gamma^\mu \partial_\mu - m_R + i\, m_I\, \gamma^5 \right ) \psi = 0~.
\end{equation}
When $i \gamma^\mu \partial_\mu$ acts on the spinor function $ \xi_r \otimes \chi_r~$, it becomes
\begin{equation}
  (i\, \sigma_3\, \partial_\tau \otimes I_{\bf 2} - i\, \sigma_2 \otimes ( {\bf k}\cdot \vec{\sigma} ))(\xi\otimes\chi)~. 
\end{equation}
Using the helicity basis relation \(({\bf k}\cdot \vec{\sigma})\chi_r=rk\chi_r\), the equation of motion of $\xi_r \otimes \chi_r$ is given by
\begin{equation}
  \Big [  \left ( i \sigma_3\, \partial_\tau - i\, r k\, \sigma_2 - m_R\, I_{\bf 2} + i\, m_I\, \sigma_1  \right ) \otimes I_{\bf 2} \Big ] \left ( \xi_r \otimes \chi_r \right ) = 0~,
\end{equation}
from which the first-order differential equation for $\xi_r$ is derived:
\begin{equation}\label{eq:xi:eom:generic}
\begin{split}
\partial_\tau\, \xi_r = - i\, \left ( {\bf q}\cdot {\vec\sigma} \right ) \xi_r~,
\end{split}
\end{equation}
where 
\begin{equation}\label{eq:qvec:psi}
   {\bf q} = rk\, \hat{x}_1+ m_I\, \hat{x}_2 + m_R\, \hat{x}_3~.
\end{equation}
We emphasize that the equation of motion for $\xi_r$ in Eq.~(\ref{eq:xi:eom:generic}) is universal in that its form is valid for any system (and for any choice of basis), and all the information about the given system (or choice of basis) are encoded in the $SO(3)_1$ vector $\bf{q}$. The ${\bf q}\cdot {\vec\sigma}$ is an embedding of $SO(3)_1$ vector ${\bf q}$ into the $SU(2)_1$ representation. 
The differential equation written in terms of $SU(2)_1$ representation in Eq.~(\ref{eq:xi:eom:generic}) can be converted into the form in terms of $SO(3)_1$ representation. The differentiation of $\vec{\zeta}_r$ with respect to time, using Eq.~(\ref{eq:xi:eom:generic}), gives rise to
\begin{equation}\label{eq:eom:zeta:xi:generic}
 \partial_\tau \zeta_{r\, i} = \frac{1}{2} \xi_r^\dagger \left [i\, {\bf q}\cdot \vec{\sigma},\,  \sigma_i \right ] \xi_r= 2\, \epsilon_{ijk}\, q_j \zeta_{r\, k}~.
\end{equation}
The equation of motion of $\vec{\zeta}_r$ is given in the vector form,
\begin{equation}\label{eq:zeta:eom:generic}
\frac{1}{2} \partial_\tau \vec{\zeta}_r = {\bf q}\times \vec{\zeta}_r~.
\end{equation}
Remarkably, the above differential equation in Eq.~(\ref{eq:zeta:eom:generic}) is nothing but the equation of motion for a vector (${\bf r}$) precessing with an angular velocity (${\vec\omega}_{\bf r}$), namely $d{{\bf r}}/d\tau = \vec{\omega}_{{\bf r}} \times {\bf r}$. As was mentioned before, choosing any interaction type of interest or any particular basis simply amounts to selecting the corresponding vector ${\bf q}$ which can be interpreted as an angular velocity of $\vec{\zeta}_r$ in the classical system.

In order to derive the fermion production in the inertial frame, we need to quantize the Dirac fermion $\psi$ in the Hamiltonian for the fermions,
\begin{equation}
\begin{split}
 \mathcal{H} &= \bar{\psi} \left ( - i \gamma^i \partial_i + m_R - i\, m_I\, \gamma^5 \right ) \psi~.
\end{split}
\end{equation}
With the expression of the quantum field $\psi$ in Eq.~(\ref{eq:psi:quantum}), we obtain the Hamiltonian in terms of the creation and annihilation operators,
\begin{equation}\label{eq:Eenergy:operators}
  {\mathcal{H}} = \sum_{r=\pm}\int d^3k \left ( a^\dagger_r ({\bf k}),\, b_r (-{\bf k})  \right ) \begin{pmatrix} A_r & B^*_r \\[2pt] B_r & - A_r\end{pmatrix} \begin{pmatrix} a_r ({\bf k}) \\[2pt]  b^\dagger_r (-{\bf k})  \end{pmatrix}~,
\end{equation}
where the matrix element is given by~\footnote{The fermion production can be considered to be analogous to the precession of the magnetic dipole (or magnetization) around the magnetic field with the angular velocity, $\vec{\omega}_{\bf M} = - \gamma {\bf B}$, whose dynamics is governed by $d{\bf M}/d\tau = \vec{\omega}_{\bf M}  \times {\bf M}$ (known as Bloch equation). The energy of the classical system, $\vec{\omega}_{\bf M} \cdot {\bf M}$, is analogous to ${\bf q}\cdot \vec{\zeta}_r$, which appears as the diagonal element of the Hamiltonian, and it becomes the energy eigenvalue of the Hamiltonian due to the vanishing off-diagonal elements when $\vec{\zeta}_r$ is parallel or anti-parallel to the ${\bf q}$ vector.}
\begin{equation}\label{eq:ArBr:H:inertial}
\begin{split}
   A_r = {\bf q}\cdot \vec{\zeta}_r~,\quad
   |B_r|^2  =  ( {\bf q}\times \vec{\zeta}_r )^2~.
\end{split}
\end{equation}
The second relation in Eq.~(\ref{eq:ArBr:H:inertial}) is nothing but the eigenvalue equation whose energy eigenvalues are $\pm |{\bf q}|$ (see Eq.~(\ref{eq:app:Ar:Br}) in Appendix~\ref{sec:app:particlenumber} for the explicit expression of $B_r$ up to the phase). One notes that the inner product ${\bf q}\cdot \vec{\zeta}_r$ is invariant under $SU(2)_1\times U(1)$. Although the system starts with the diagonalized Hamiltonian, the Hamiltonian after a time $t$ generally becomes non-diagonal, and therefore, the operators $a^\dagger_r$ and $b^\dagger_r$ (and $a_r$ and $b_r$) at a later time $\tau$ do not create (and destroy) energy eigenstates. The creation and annihilation operator after diagonalizing the Hamiltonian becomes an admixture of the operators before the diagonalization, and they gain the time-dependence through $u_r$ and $v_r$ functions. Expressing the Hamiltonian in Eq.~(\ref{eq:Eenergy:operators}) in terms of the creation and annihilation operators, which correspond to the one-particle states, amounts to
\begin{equation}\label{eq:rotation}
 \begin{pmatrix} a_r ({\bf k}) \\[2pt]  b^\dagger_r (-{\bf k})  \end{pmatrix} 
 \rightarrow 
 \begin{pmatrix} \alpha^*_r & \beta^*_r \\[2pt] -\beta_r & \alpha_r\end{pmatrix}
 \begin{pmatrix} a_r ({\bf k}) \\[2pt]  b^\dagger_r (-{\bf k})  \end{pmatrix}~,
\end{equation}
where mixing angles, $\alpha_r$ and $\beta_r$, are called Bogoliubov coefficients, and they are linear in $u_r$ and $v_r$ as the matrix elements in Eq.~(\ref{eq:ArBr:H:inertial}) are linear in $\vec{\zeta}_r$ (or quadratic in $u_r$ and $v_r$).

The particle number (similarly for anti-particle) for a helicity $r$ is defined as
\begin{equation}\label{eq:N:define}
  N_r(\tau) = \langle 0| \int \frac{d^3 k}{(2\pi)^3}\, a_r^\dagger\, a_r | 0\rangle  \equiv \int d^3 k\, n_{r,\, k} (\tau)~,
\end{equation}
where $n_{r,\, k}(\tau)$ is the particle number density for a $k$ mode, and the operators $a_r^\dagger$ and $a_r$ are associated with the one-particle state at time $\tau$. From the point of view of the time-varying creation and annihilation operators, the vacuum $|0\rangle$ in Eq.~(\ref{eq:N:define}) is the one defined at the initial time where the Hamiltonian takes a diagonal form, or the particle number density is initially zero as it should. 
Due to the Pauli exclusion principle, it must be always smaller than (or equal to) unit,
\begin{equation}\label{eq:nk:pauliblocking}
 0 \leq n_{r,\, k}(\tau) = |\beta_r|^2 \leq 1~,
\end{equation}
and it is known as the Pauli-blocking. 
While the analytic expression of $n_{r,\, k}(\tau)$ is obtained by a complicated algebra in the traditional approach, its expression can be uniquely determined by a few properties in our group theoretic approach. As was explained in Section~\ref{sec:grouptheoryapproach}, the length of $\vec{\zeta}_r$ is unit, or $|\vec{\zeta}_r| = 1$. Since the length of the $\vec{\zeta}_r$ is preserved, the inequality,
\begin{equation}\label{eq:inequality:stepI}
-1 \leq \frac{{\bf q} \cdot \vec{\zeta}_r}{|{\bf q}|} = \cos \theta \leq 1~,
\end{equation}
holds over the time evolution. The $\theta$ is the angle between two vectors $\vec{\zeta}_r$ and ${\bf q}$. Since the mixing angle in the diagonalization of the Hamiltonian should be linear in $u_r$ and $v_r$, the particle number density in Eq.~(\ref{eq:N:define}) can be at most a linear function in $\vec{\zeta}_r$ (or quadratic in $u_r$ and $v_r$). It can be written as, up to a sign ambiguity,
\begin{equation}\label{eq:nk:ansatz}
 n_{r,\, k}(\tau) = A \pm B\, \frac{{\bf q} \cdot \vec{\zeta}_r}{|{\bf q}|}~.
\end{equation}
The negative sign in front of $B$ in Eq.~(\ref{eq:nk:ansatz}) has to be chosen to be consistent with the form of the Hamiltonian in Eqs.~(\ref{eq:Eenergy:operators}) and~(\ref{eq:ArBr:H:inertial}). For instance, when $\vec{\zeta}_r$ is parallel to the ${\bf q}$ vector, ${\bf q} \cdot \vec{\zeta}_r$ corresponds to the energy eigenvalue due to the vanishing off-diagonal matrix elements. In this situation, the $a_r$ and $a_r^\dagger$ operators in Eq.~(\ref{eq:N:define}) correspond to the one-particle states which leads to the zero particle number density (see Appendix~\ref{sec:app:particlenumber} for a detailed discussion).

The inequality in Eq.~(\ref{eq:inequality:stepI}) implies that
\begin{equation}\label{eq:inequality:stepII}
  A - B \leq n_{r,\, k}(\tau) \leq A + B~.
\end{equation}
Matching both sides of Eq.~(\ref{eq:inequality:stepII}) to those in Eq.~(\ref{eq:nk:pauliblocking}) determines two coefficients, $A$ and $B$, and gives rise to the analytic expression of the particle number density (see Appendix~\ref{sec:app:particlenumber} for an explicit derivation),
\begin{equation}\label{eq:nk:zeta:generic}
\begin{split}
 n_{r,\, k} (\tau) =&~ \frac{1}{2}\left ( 1 - \frac{{\bf q} \cdot \vec{\zeta}_r}{|{\bf q}|} \right ) = \frac{1}{2} \left ( 1 - \cos \theta \right )~.
\end{split}
\end{equation}
An advantage of the form in Eq.~(\ref{eq:zeta:eom:generic}) in terms of $\vec{\zeta}_r$, compared to Eq.~(\ref{eq:xi:eom:generic}) in terms of $\xi_r$, is that the correct initial condition for the $\vec{\zeta}_r$ corresponding to zero particle production at $\tau=\tau_0$ is straightforward in Eq.~(\ref{eq:nk:zeta:generic}) which is
\begin{equation}\label{eq:zeta:bc}
 \vec{\zeta}_r(\tau_0,\, \tau_0) = \frac{{\bf q}(\tau_0)}{|{\bf q}(\tau_0)|} \equiv \frac{{\bf q}_0}{|{\bf q}_0|}~.
\end{equation}
From the point of view of the explicit derivation of the fermion number density in Eq.~(\ref{eq:nk:zeta:generic}) (as was done in Appendix~\ref{sec:app:particlenumber}), our group theoretic formalism explicitly shows the feature of the Pauli-blocking, namely $0\leq n_{r,\, k} (\tau) \leq 1$, in terms of an angle between two vectors, ${\bf q}$ and $\vec{\zeta}_r$, with which one can visualize the fermion production dynamics.

Just like solving the Schr$\ddot{\rm o}$dinger equation for the unitary operator in quantum mechanics, the closed form of the solution for $\vec{\zeta}_r$ can be easily obtained. We rewrite the Eq.~(\ref{eq:zeta:eom:generic}) in a matrix form,
\begin{equation}
\frac{\partial \vec{\zeta}_r (\tau,\, \tau_0)}{\partial \tau} = M(\tau)\, \vec{\zeta}_r (\tau,\, \tau_0) \quad \hbox{with} \quad \vec{\zeta}_r(\tau_0,\, \tau_0) = \frac{{\bf q}_0}{|{\bf q}_0|}~,
\end{equation}
where the matrix $M(\tau)$ can be written as $M(\tau)= {\bf q}\cdot {{\bf L}}$ with ${{\bf L}}$ being the $3\times 3$ matrix representation of the $SO(3)_1$ group and ${\bf q}$ is the vector in Eq.~(\ref{eq:qvec:psi}). We can solve the differential equation iteratively order-by-order in $M(\tau)$. The final solution is given by
\begin{equation}\label{eq:zeta:exact:Tordering:generic}
\begin{split}
 \vec{\zeta}_r (\tau,\, \tau_0) =  T\, \exp{ \left ( \int_{\tau_0}^\tau d\tau'\, M(\tau') \right ) }\, \frac{{\bf q}_0}{|{\bf q}_0|}~,
\end{split}
\end{equation}
where $T$ denotes a time-ordering.   Finally, the resulting particle number density is given by
\begin{equation}\label{eq:nk:psi:final}
 n_{r,\, k} (\tau) = \frac{1}{2}\left ( 1 - \frac{{\bf q}(\tau)}{|{\bf q}(\tau)|}\cdot T\, \exp{ \left ( \int_{\tau_0}^\tau d\tau'\, M(\tau') \right ) }\, \frac{{\bf q}_0}{|{\bf q}_0|} \right )~.
\end{equation}
While the form in Eq.~(\ref{eq:nk:psi:final}) takes a closed form, it is a separate issue whether it is practically useful or not unless one can extract any type of (semi) analytic expression out of it. Since $M(\tau)= {\bf q}\cdot {{\bf L}}$ and the matrices $L_i$ satisfies the commutation relation, one might expect that the expression in Eq.~(\ref{eq:nk:psi:final}) can be further processed to obtain an analytic expression. However, we have not managed to simplify the solution.

We close this section by comparing our result with literature. By plugging Eqs.~(\ref{eq:zeta:comp}) and~(\ref{eq:qvec:psi}) into Eq.~(\ref{eq:nk:zeta:generic}) and defining $|{\bf q}| \equiv \omega$, the particle number density in terms of $u_r$ and $v_r$ is given by
\begin{equation}\label{eq:nk:urvr}
  n_{r,\, k} (\tau) 
  = \frac{1}{2}-\frac{m_R}{4\omega} \left ( |u_r|^2-|v_r|^2 \right ) - \frac{k}{2\omega} Re(u_r^* v_r) - \frac{r\, m_I}{2\omega} Im (u_r^* v_r)~,
\end{equation}
and this expression agrees with the result in~\cite{Adshead:2018oaa}. An agreement with those in~\cite{Adshead:2018oaa} is also hold for the expressions for $A_r$ and $|B_r|^2$ in Eq.~(\ref{eq:ArBr:H:inertial}) in terms $u_r$ and $v_r$ (see Eqs.~(\ref{eq:app:Ar:Br}) and~(\ref{eq:app:vanishingBr}) for the explicit derivation).

\subsection{Fermion Production in Rotating Frame}
\label{sec:Fermion:rotatingframe}
The transformation of the Lagrangian in Eq.~(\ref{eq:Lag:psi}) to the one in Eq.~(\ref{eq:Lag:Y}) with the derivative coupling of the pseudo-scalar to the fermions,
\begin{equation}\label{eq:rot:psitoY}
  \psi \rightarrow e^{i \gamma^5 \phi/f} \psi~,
\end{equation}
amounts to the $\phi(\tau)$-dependent $SO(3)_1$ rotation, $\vec{\zeta}_r \rightarrow R(\tau)\, \vec{\zeta}_r$ where the time-dependent rotation matrix $R(\tau)$ is given by
\begin{equation}\label{eq:SO3:rot:matrix}
   R(\tau) = \begin{pmatrix} 1 & 0 & 0 
   \\[3pt] 
   0 & \cos  \frac{2\phi}{f} & -\sin \frac{2\phi}{f}
   \\[3pt] 
   0 & \sin  \frac{2\phi}{f}  & \cos \frac{2\phi}{f}
    \end{pmatrix}~,
\end{equation}
and it corresponds to the rotation by $2\phi/f$ angle around the $\zeta_{r\, 1}$ axis. We will call the transformed $\vec{\zeta}_r$ frame a  rotating frame to distinguish it from the inertial $\vec{\zeta}_r$ frame in Section~\ref{sec:Fermion:Inertialframe}.

Moving into the rotating frame via the time-dependent rotation in classical mechanics introduces fictitious forces, which have no physical origin, such as the coriolis force, centrifugal force, and a term related to the acceleration of the axes. Those fictitious forces need to be introduced in the rotating frame to make the physics frame-independent. Following the analogy to the classical mechanics, we would expect similar fictitious terms to be introduced when moving into the rotating $\vec{\zeta}_r$ frame, or basis with the derivative coupling of the pseudo-scalar to fermions, via the time-dependent rotation with the matrix in Eq.~(\ref{eq:SO3:rot:matrix}). 

Under the time-dependent rotation with the matrix $R(\tau)$, the equation of motion for $\vec{\zeta}_r$ transforms like
\begin{equation}
  \frac{1}{2} \partial_\tau \vec{\zeta}_r = \left ( {\bf q} \cdot {{\bf L}} \right )  \vec{\zeta}_r~\quad
  \rightarrow \quad
    \frac{1}{2} \partial_\tau \left ( R\, \vec{\zeta}_r \right ) =  \left ( {\bf q} \cdot {{\bf L}} \right ) R\, \vec{\zeta}_r~.
\end{equation}
The equation of motion in the rotating frame can be written as~\footnote{The rotation matrix $R$ for an orthogonal group satisfies $R^T R = 1$. Differentiating the relation with respect to time gives $\dot{R}^T R + R^T \dot{R} = (R^T \dot{R})^T + R^T \dot{R}= 0$ which implies that $R^T \dot{R}$ is antisymmetric. We can define a vector $\vec{\omega}_{\zeta_r}$ such that $(R^T \dot{R})_{ij} \equiv \epsilon_{ijk}\omega_{\zeta_r\, k}$. Therefore, $- R^T \partial_\tau R\, \vec{\zeta}_r = \vec{\omega}_{\zeta_r} \times \vec{\zeta}_r$.}
\begin{equation}
  \frac{1}{2} \partial_\tau\, \vec{\zeta}_r =  \left ( R\, {\bf q} \right ) \cdot {\bf L}\, \vec{\zeta}_r + \frac{1}{2}\, \vec{\omega}_{\zeta_r}\times \vec{\zeta}_r~,
\end{equation}
where we used $R^T \left ( {\bf q} \cdot {\bf L} \right ) R = (R\, {\bf q})\cdot {\bf L}$, and the $\vec{\omega}_{\zeta_r}$ can be interpreted as the angular velocity of the rotating $\vec{\zeta}_r$ axes which is given by
\begin{equation}
  \vec{\omega}_{\zeta_r} = \begin{pmatrix} 2 \dot{\phi}/f \\[3pt] 0 \\[3pt] 0 \end{pmatrix}~.
\end{equation}
When the equation of motion for the transformed $\vec{\zeta}_r$ is brought back into the universal form,
\begin{equation}\label{eq:zeta:rotframe}
  \frac{1}{2} \partial_\tau\, \vec{\zeta}_r = \left ( R\, {\bf q} + \frac{1}{2}\, \vec{\omega}_{\zeta_r} \right ) \times \vec{\zeta}_r =  \tilde{\bf q}\times \vec{\zeta}_r~,
\end{equation}
the $\tilde{\bf q}$ in the rotating frame is obtained by
\begin{equation}\label{eq:qprime:rotating}
  \tilde{\bf q} = \left ( r k + \frac{\dot{\phi}}{f} \right )\, \hat{x}_1 + ma\, \hat{x}_3~.
\end{equation}
As is evident in Eqs.~(\ref{eq:zeta:rotframe}) and~(\ref{eq:qprime:rotating}), the differential equation for $\vec{\zeta}_r$ stays in a universal form, and the information on the rotating frame is encoded in the new $\tilde{\bf q}$ vector.
The equation of motion in Eq.~(\ref{eq:zeta:rotframe}) agrees with the one derived directly from the Dirac equation from the Lagranagian in Eq.~(\ref{eq:Lag:Y}),
\begin{equation}
  \Big [ \Big ( i\, \sigma_3 \partial_\tau - i\, rk \sigma_2 - ma\, I_{\bf 2} - i\, \frac{\dot\phi}{f} \sigma_2 \Big ) \otimes I_{\bf 2} \Big ] \left ( \xi_r \otimes \chi_r \right )= 0~,
\end{equation}
which induces the differential equation for the $SU(2)_1$ doublet $\xi_r$,
\begin{equation}
\begin{split}
\partial_\tau\, \xi_r = - i\, \left ( \tilde{\bf q}\cdot {\vec\sigma} \right ) \xi_r~,
\end{split}
\end{equation}
where $\tilde{\bf q}$ is the same as Eq.~(\ref{eq:qprime:rotating}).

Since the $\tilde{\bf q}$ and $\vec{\zeta}_r$ are the only available $SO(3)_1$ vectors, the particle number density for $k$ mode in the rotating $\vec{\zeta}_r$ frame needs to be a function of the inner product, $\tilde{\bf q} \cdot \vec{\zeta}_r$ (and lengths of $\tilde{\bf q}$ and $\vec{\zeta}_r$),
\begin{equation}\label{eq:nk:zeta:generic:rotating}
  n_{r,\, k}(\tau) = f \left ( \tilde{\bf q} \cdot \vec{\zeta}_r \right )~.
\end{equation}

While the particle number density in Eq.~(\ref{eq:nk:zeta:generic}) is conserved under a time-independent $SO(3)_1$ rotation, which can be thought of changing from an inertial $\vec{\zeta}_r$ frame to another inertial $\vec{\zeta}_r$ frame, we suspect that the particle number density~\footnote{What we meant by the particle number density here is the one defined based on the first principle, namely taking the quadratic part in the Hamiltonian, diagonalizing it, and defining the particle number as the expectation value of the number operator.} changes in the transition from the inertial frame to the rotating frame, or non-inertial frame. The particle number density must be at most linear in $\vec{\zeta}_r$ in the rotating frame as well, and it is similarly determined to be~\footnote{The result in Eq.~(\ref{eq:nk:zeta:rotating}) matches to the $|\tilde{\beta}_r|^2$ in the Appendix B of~\cite{Adshead:2018oaa}, which is a particle number density obtained from the quadratic term in $\psi$ in Eq.~(\ref{eq:Y:hamiltonian}).}
\begin{equation}\label{eq:nk:zeta:rotating}
 n_{r,\, k} (\tau) = \frac{1}{2}\left ( 1 - \frac{\tilde{\bf q} \cdot \vec{\zeta}_r}{|\tilde{\bf q}|} \right )~.
\end{equation}
All higher-order terms in $\vec{\zeta}_r$ should be forbidden by demanding that the particle number densities in two frames should match in the $\dot\phi \rightarrow 0$ limit (time-independent rotation limit). One notes that the particle number in Eq.~(\ref{eq:nk:zeta:rotating}) matches to the one that is derived from the free Hamiltonian following similar steps to Appendix~\ref{sec:app:particlenumber}.

As was explained in Section~\ref{sec:model}, the Hamiltonian in the basis with the derivative coupling (or in the rotating frame, or the non-inertial frame, in our language) does not take a simple quadratic form in $\psi$ with an obvious decoupling limit when the fermion mass vanishes, and the velocity of $\phi$, including the fermion bilinear term, introduces the fermion quartic coupling. This complication prevents us from estimating the final particle number density at a later time unambiguously in the rotating $\vec{\zeta}_r$ frame, whereas the particle number density can be unambiguously estimated in the inertial frame. However, one should note that the nature will not care about the choice of the basis, or the preference of the inertial $\vec{\zeta}_r$ frame is not a physical consequence.

\subsection{Backreaction due to Fermion Production}
\label{sec:backreaction}
In this section, we will briefly discuss about the backreaction of the produced fermion on the pseudo-scalar dynamics. 
The zero backreaction in the massless limit can be reinterpreted in our group theoretic formalism. We take the equation of motion of $\phi$ from the Lagrangian in Eq.~(\ref{eq:Lag:psi}),
\begin{equation}
  \ddot{\phi} + 2 \frac{\dot{a}}{a} \dot\phi + a^2 V'(\phi)= \frac{2}{a^2 f} \langle  \bar{\psi} \left ( m_I + i\, m_R\, \gamma^5 \right ) \psi \rangle~.
\end{equation}
Using the simplified form of $\psi$ in a tensor product form, where the two-component spinor $\chi_r$ can be dropped due to the orthgonormality, in Appendix~\ref{sec:app:particlenumber}, we trivially obtain
\begin{equation}\label{eq:backreaction:drag:psi}
  \langle  \bar{\psi} \left ( m_I + i\, m_R\, \gamma^5 \right ) \psi \rangle 
  =
  -\sum_{r=\pm}  \int \frac{d^3 k}{(2\pi)^3}\, \langle m_I\, \zeta_{r\, 3} + m_R\, \zeta_{r\, 2} \rangle~.
\end{equation}
In the massless limit $m \rightarrow 0$, the above expression in Eq.~(\ref{eq:backreaction:drag:psi}) obviously becomes zero as it is proportional to the mass term. However, we can also see that the only non-vanishing component of $\vec{\zeta}_r$ in the massless limit is $\zeta_{r\, 1}$. In the massless limit, the vector ${\bf q}$ becomes constant staying on the $x_1$-axis all the time, ${\bf q} = rk\, \hat{x}_1$ (see Eq.~(\ref{eq:qvec:psi})). Since there must be no particles produced at the initial time, $\vec{\zeta}_r$ should be on the $x_1$-axis too to be parallel to the vector ${\bf q}$ (see Eq.~(\ref{eq:nk:zeta:generic})). As a result, the equation of motion for $\vec{\zeta}_r$ becomes trivial, or $\partial_\tau \vec{\zeta}_r = 2\, {\bf q} \times \vec{\zeta}_r = 0$ for any time, and $\vec{\zeta}_r$ stays on the $x_1$-axis forever, or $\zeta_{r\, 2} (\tau) = \zeta_{r\, 3}(\tau) = 0$.

\section{Numerical Analysis}
\label{sec:numeric:simulation}
\begin{figure}[tph]
\begin{center}
\includegraphics[width=0.314\textwidth]{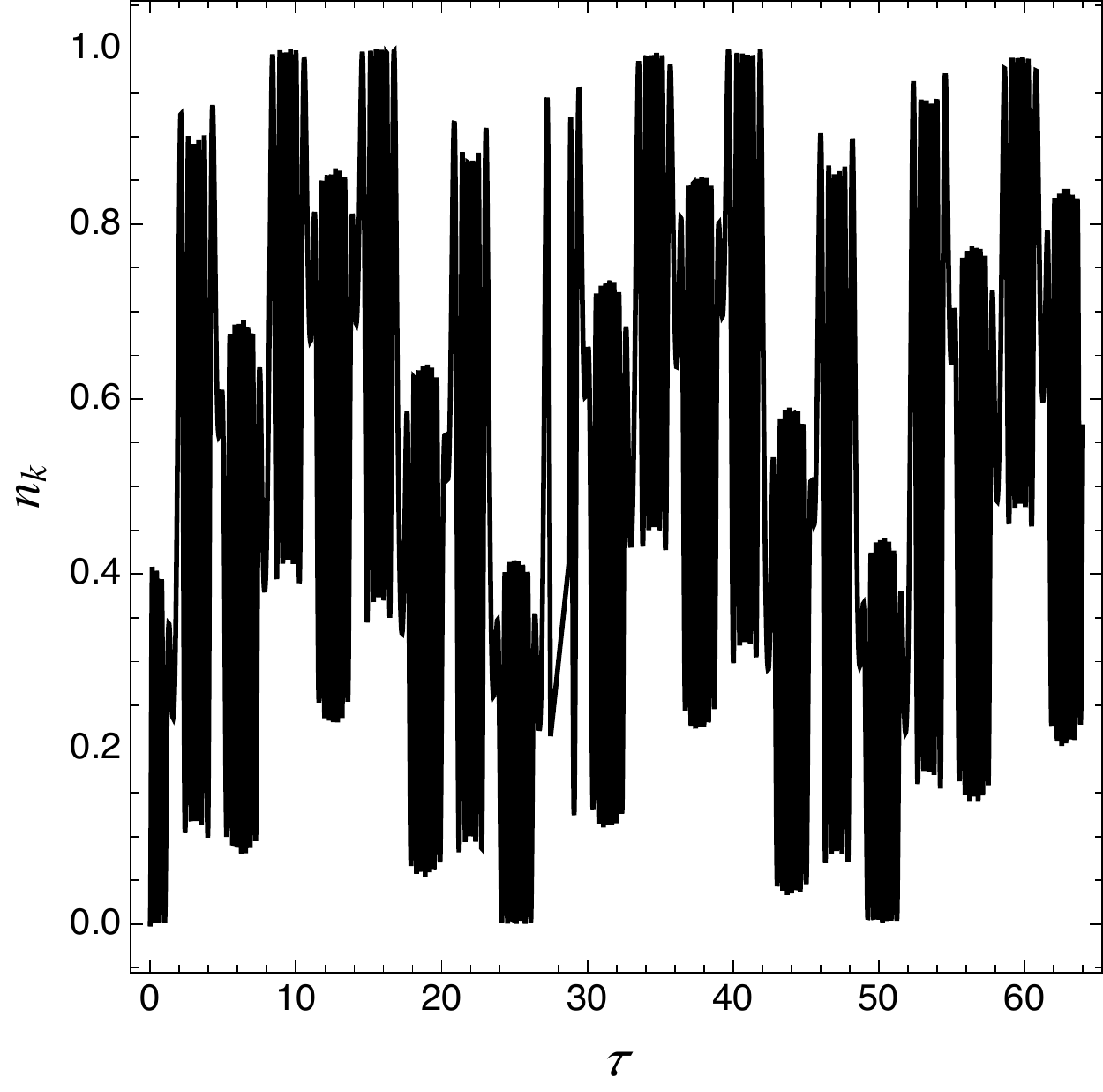}
\includegraphics[width=0.314\textwidth]{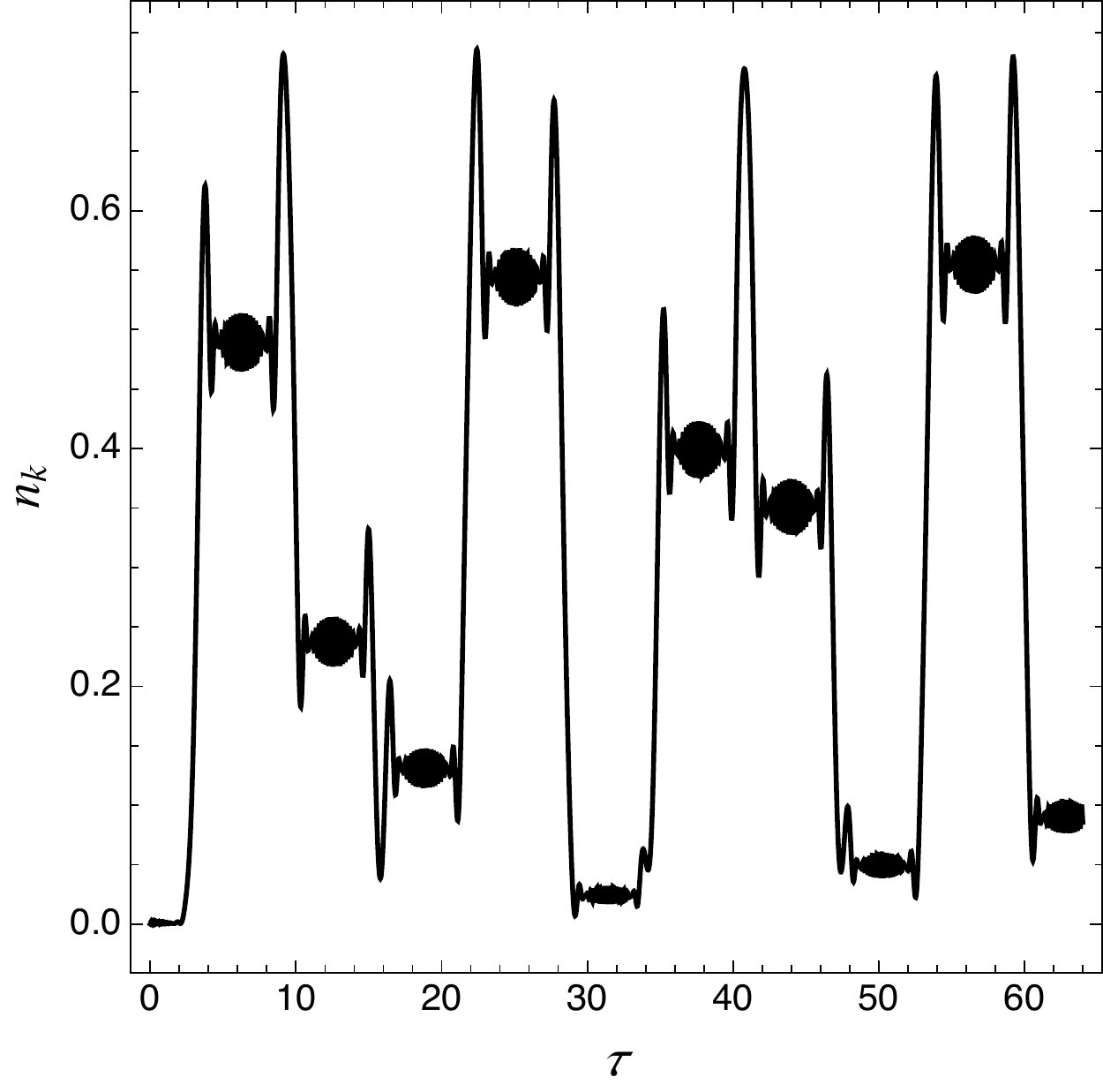}
\includegraphics[width=0.32\textwidth]{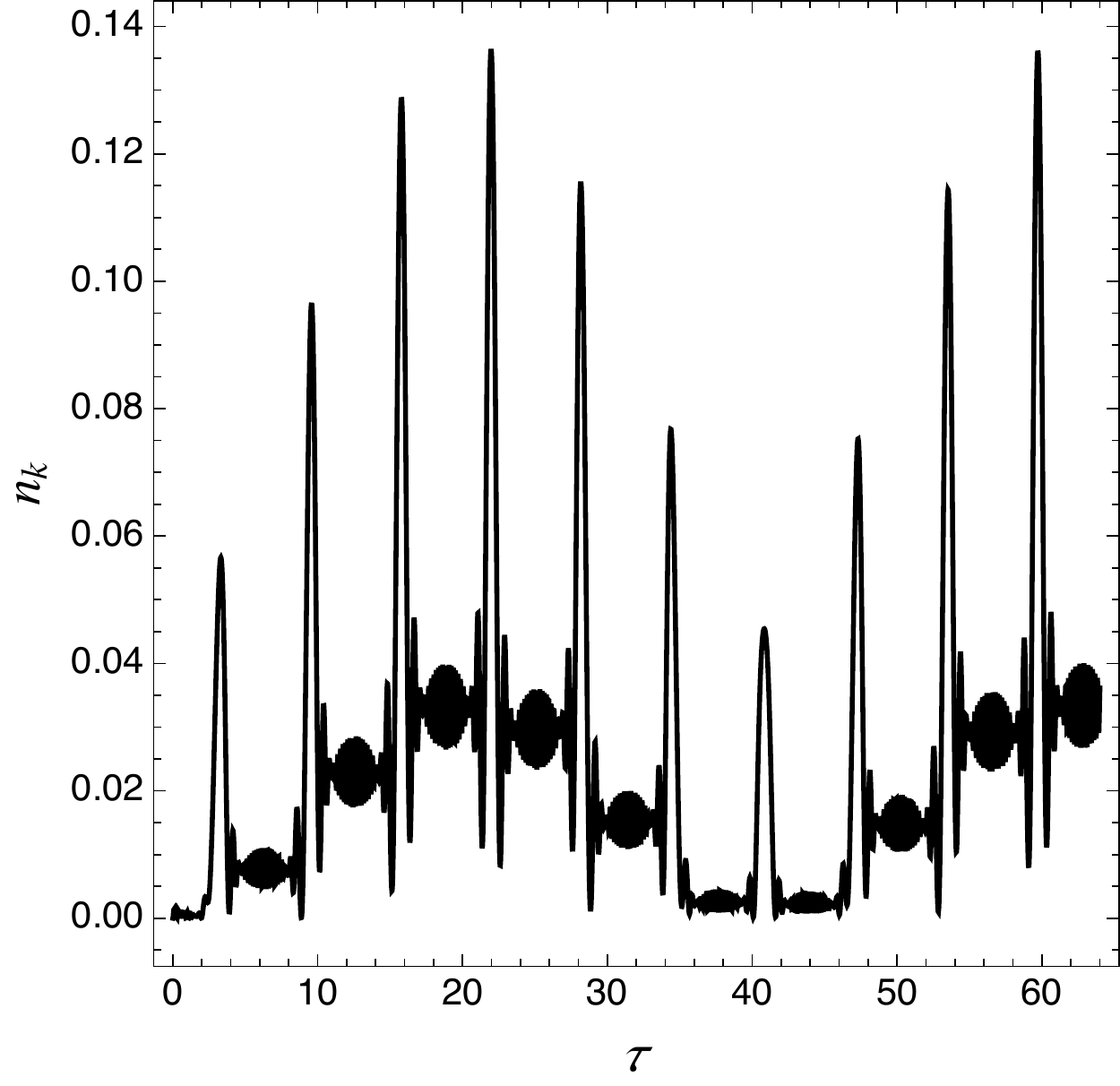}
\caption{\small The particle number density in the inertial frame for $m=1$, $\phi_0/f = 10$, and $k=1$ (leftest), $k=10$ (middle), and $k=12$ (rightest). The plots were obtained by solving the equation of motion for $\vec{\zeta}_r$ with the helicity $r=+1$ in the inertial frame.}
\label{fig:inertial:zeta:static}
\end{center}
\end{figure}

Since our group theoretic approach reproduces the same results as those from the traditional approach (as was shown in Eq.~(\ref{eq:nk:urvr})), we would expect the same numerical outcome as well. In this section, we will demonstrate how simply our new approach can simulate the fermion production compared to the traditional approach. To this end, we will reproduce some result in the literature using our method. We will also use this section to address a few subtle issues in estimating the fermion production. As a benchmark example for the illustration, we choose the following quadratic potential,
\begin{equation}
  V(\phi) = \frac{1}{2} m^2_\phi \phi^2~.
\end{equation}
In the static Universe, the solution can be parametrized as $\phi(\tau) = \phi_0 \sin(\tau)$. We numerically solved the equation for $\vec{\zeta}_r$ in Eq.~(\ref{eq:zeta:eom:generic}) in the inertial frame with the vector ${\bf q}$ in Eq.~(\ref{eq:qvec:psi}) for three same set of parameters as those in~\cite{Adshead:2015kza}, and the resulting particle number density is illustrated in Fig.~\ref{fig:inertial:zeta:static}. 
\begin{figure}[tph]
\begin{center}
\includegraphics[width=0.314\textwidth]{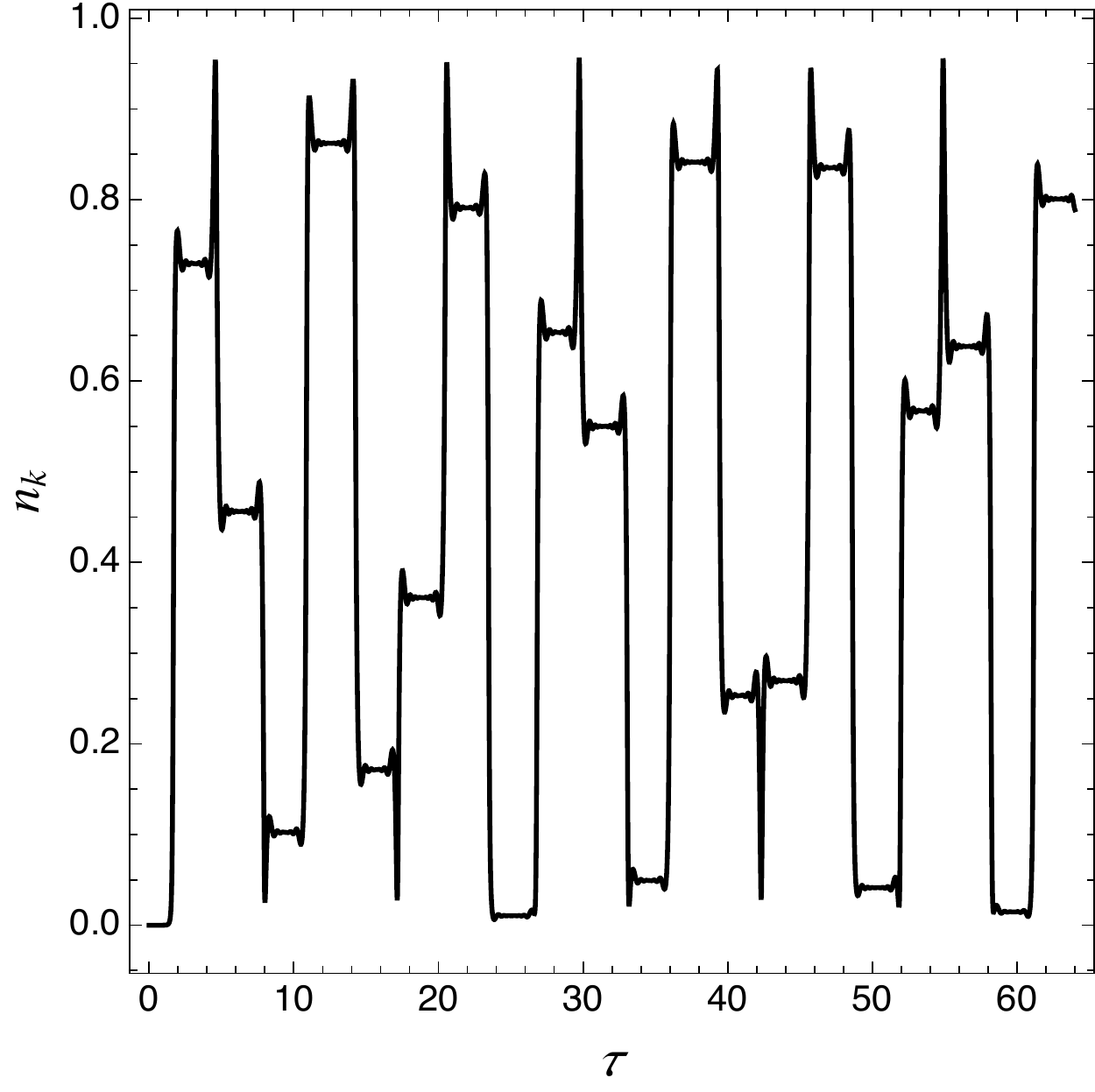}
\includegraphics[width=0.314\textwidth]{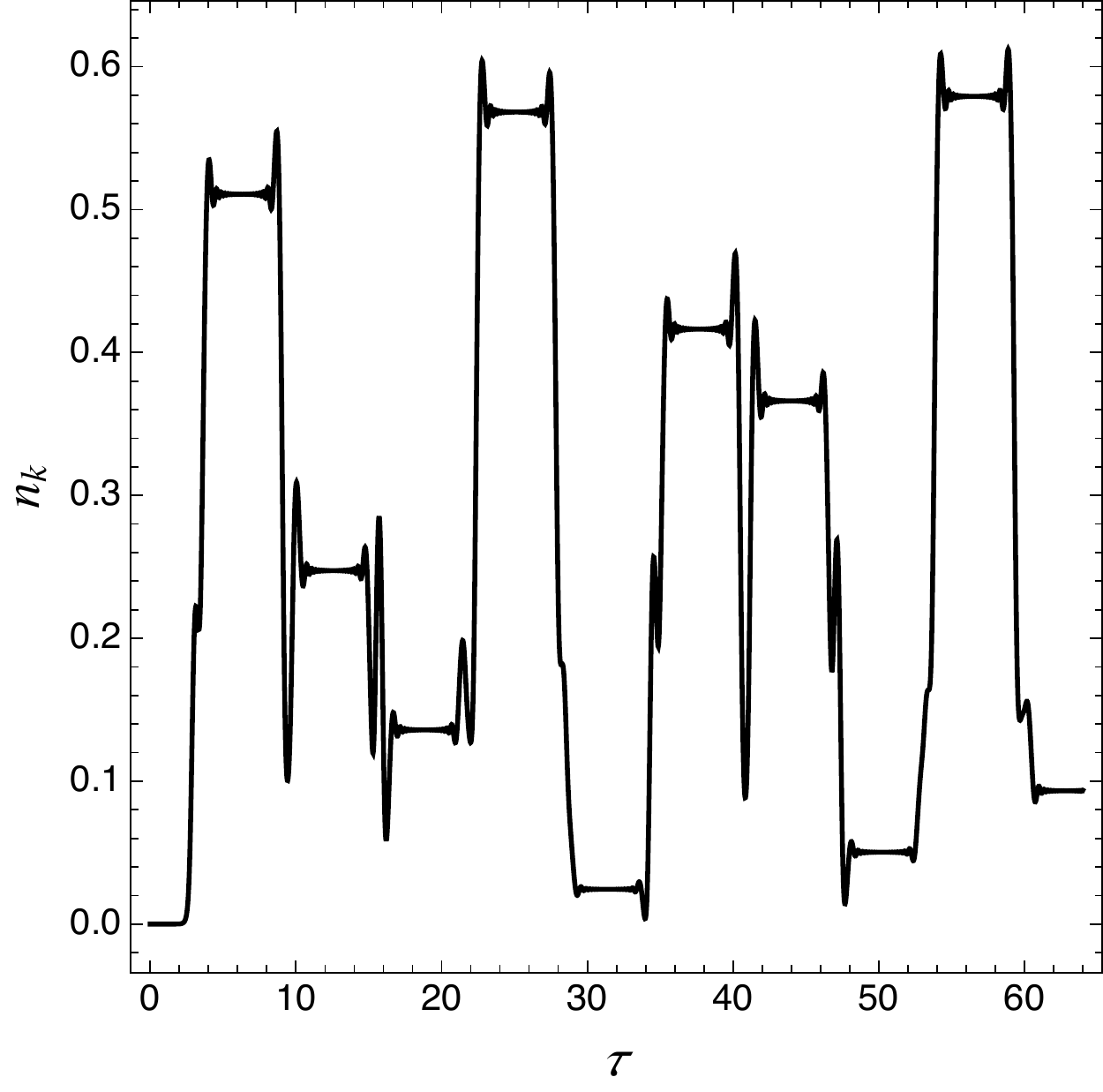}
\includegraphics[width=0.32\textwidth]{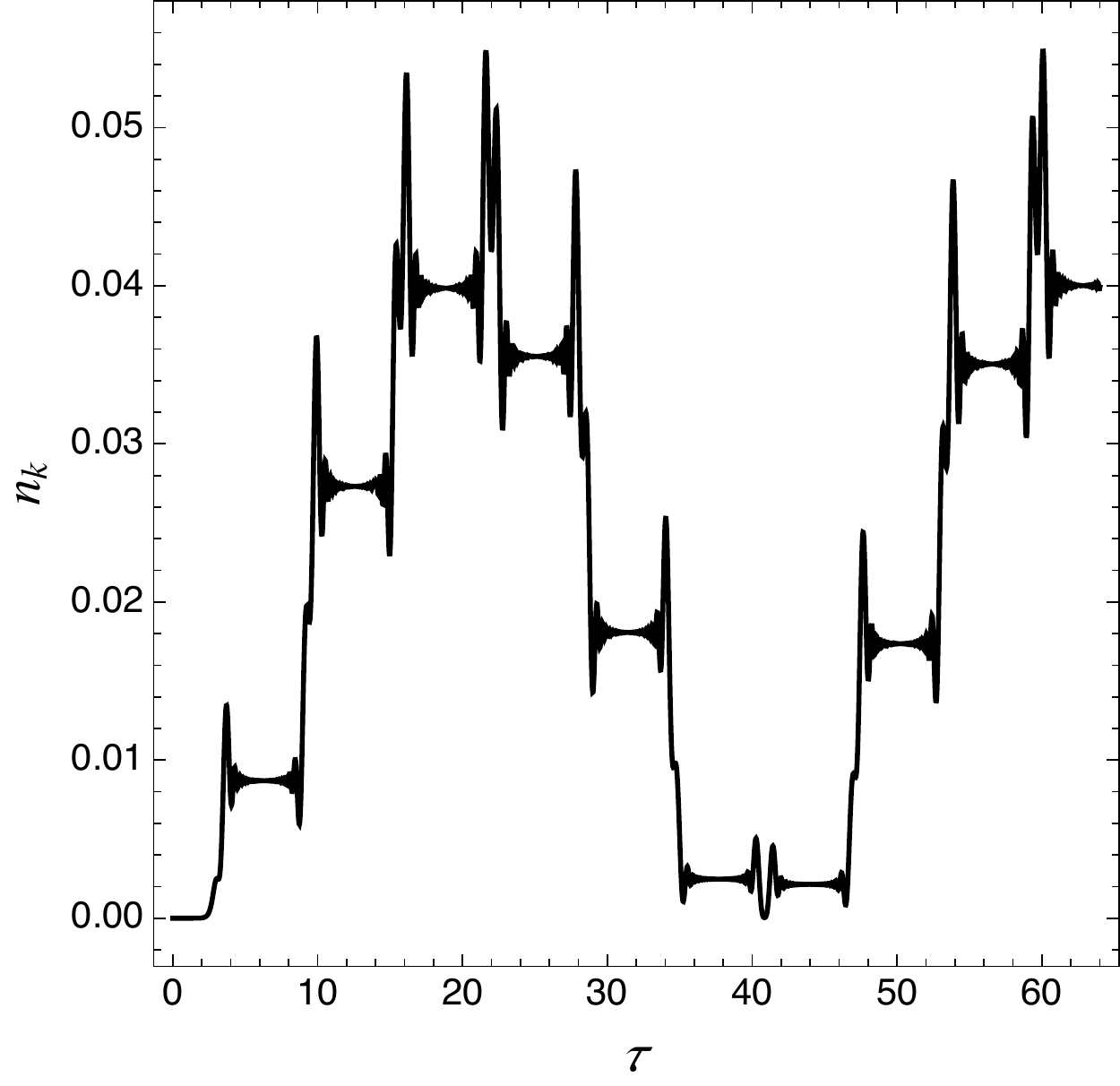}
\caption{\small The particle number density in the rotating frame for $m=1$, $\phi_0/f = 10$, and $k=1$ (leftest), $k=10$ (middle), and $k=12$ (rightest). The plots were obtained by solving the equation of motion for $\vec{\zeta}_r$ with the helicity $r=+1$ in the rotating frame.}
\label{fig:rot:zeta:static}
\end{center}
\end{figure}
Similarly particle number density in the rotating frame (although it is ambiguously defined) for the same set of parameters is shown in Fig.~\ref{fig:rot:zeta:static} where we numerically solved the equation for $\vec{\zeta}_r$ in Eq.~(\ref{eq:zeta:rotframe}) in the rotating frame with the vector $\tilde{\bf q}$ in Eq.~(\ref{eq:qprime:rotating}). The result in Fig.~\ref{fig:rot:zeta:static} exactly reproduces those in~\cite{Adshead:2015kza} (see Fig.~2 of~\cite{Adshead:2015kza}) where the particle number was estimated in the basis with the derivative coupling, or the rotating frame in our language. To make a clear comparison between two frames, we superimpose the plots in two frames and present them in Fig.~\ref{fig:both:zeta:static}. As is evident in Fig.~\ref{fig:both:zeta:static}, particle number densities in two frames are different not only in the form of expression but also numerically.
\begin{figure}[tph]
\begin{center}
\includegraphics[width=0.314\textwidth]{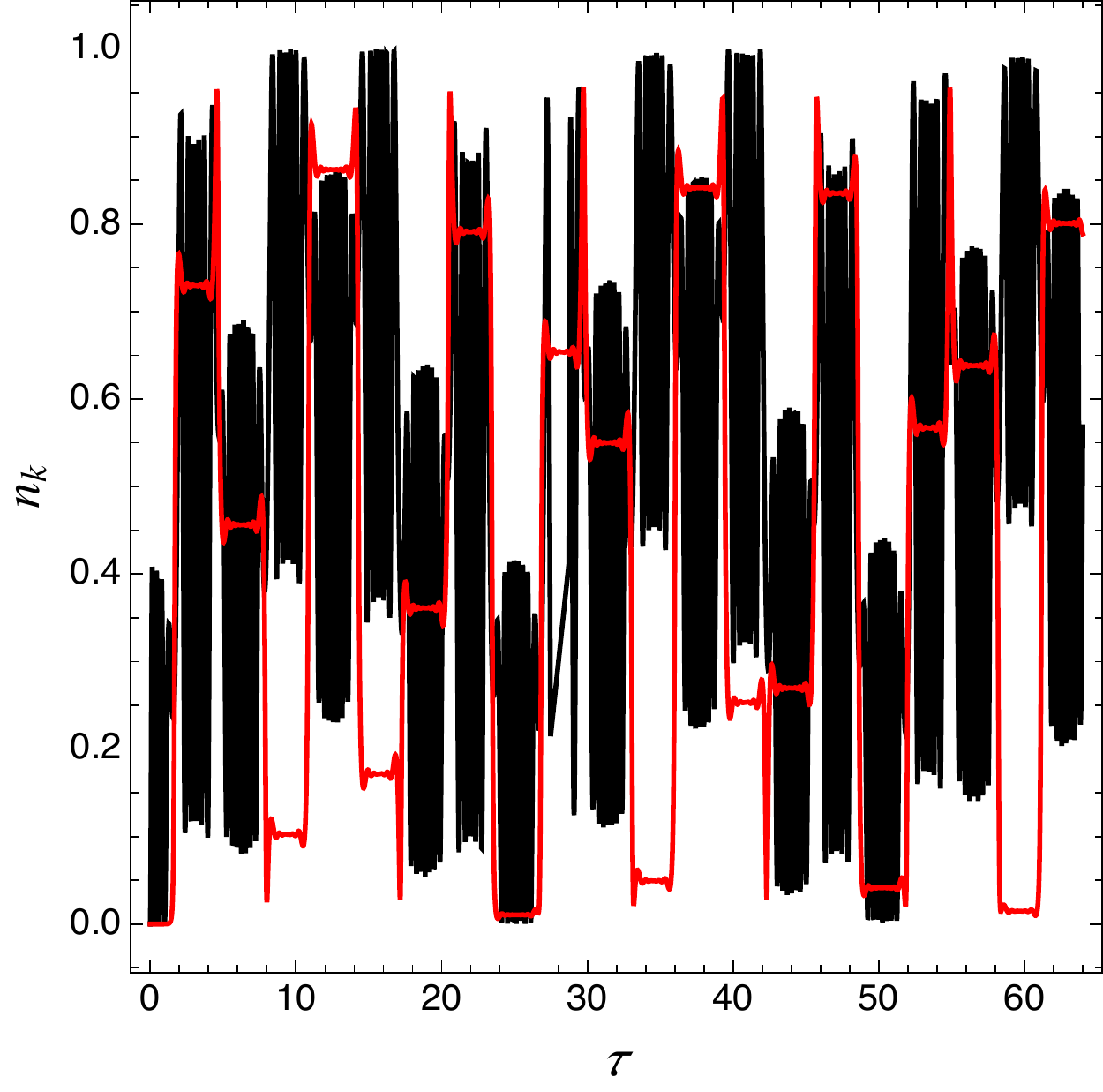}
\includegraphics[width=0.314\textwidth]{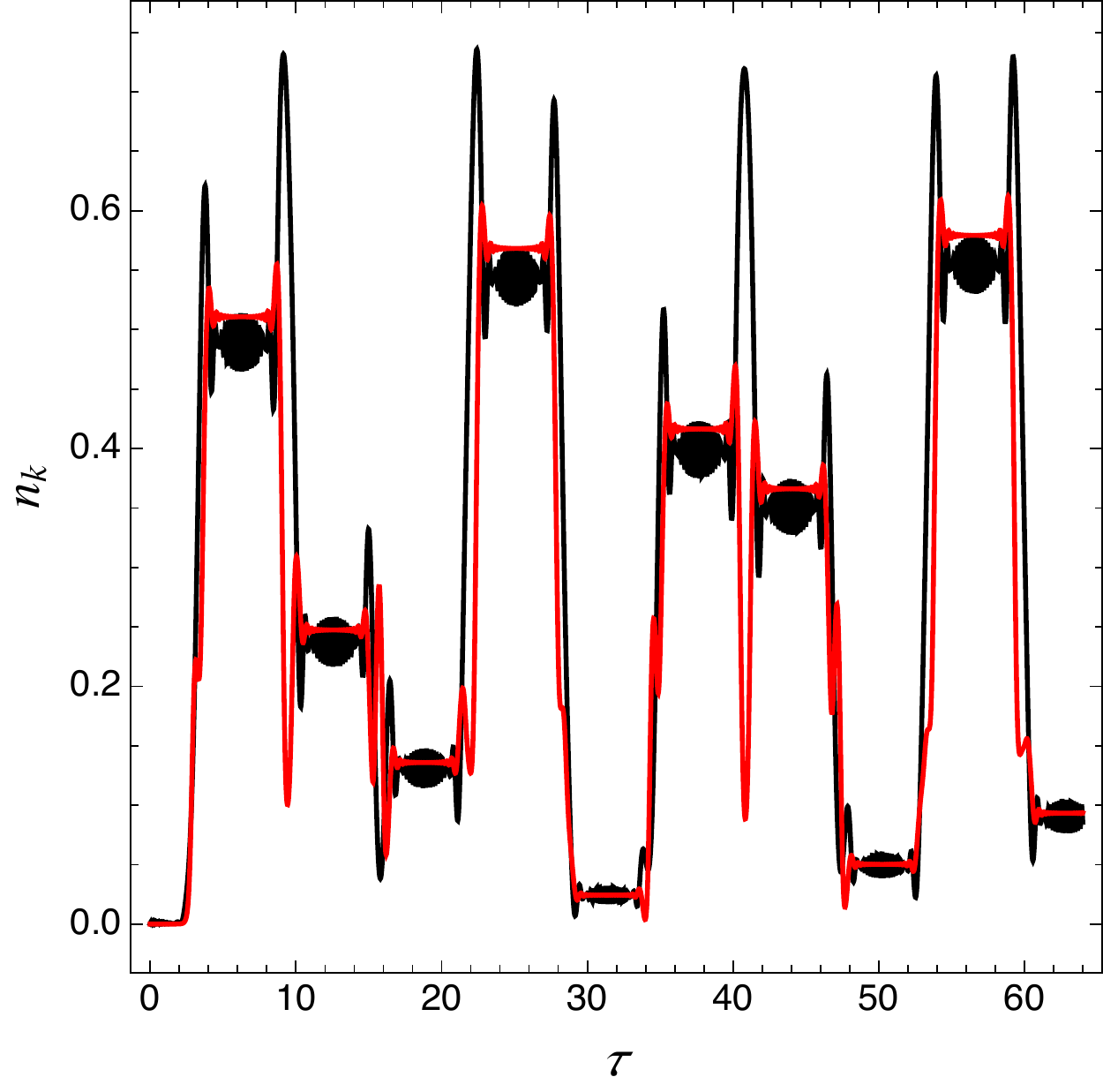}
\includegraphics[width=0.32\textwidth]{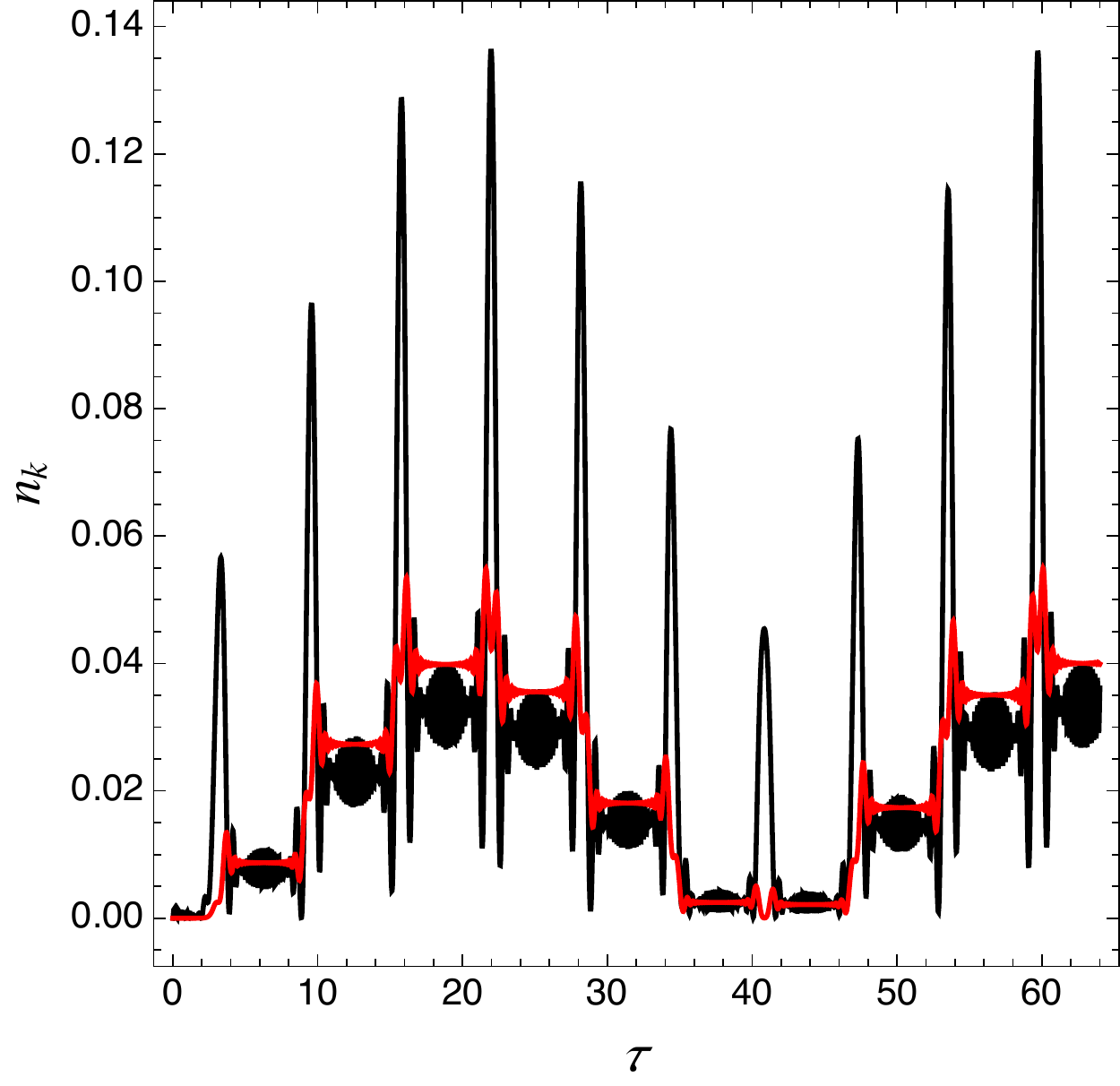}
\caption{\small The particle number density in the rotating frame (red) from Fig.~\ref{fig:rot:zeta:static} and in the inertial frame (black) from Fig.~\ref{fig:inertial:zeta:static} for $m=1$, $\phi_0/f = 10$, and $k=1$ (leftest), $k=10$ (middle), and $k=12$ (rightest).}
\label{fig:both:zeta:static}
\end{center}
\end{figure}
The discrepancy in particle number density between two frames can be better understood by looking at the explicit expression in terms of input parameters and components of the spinor function. For instance, the particle number density in the rotating frame in Eq.~(\ref{eq:nk:zeta:rotating}) can be expanded in terms of $\tilde{u}_r$ and $\tilde{v}_r$ (tilde symbol to refer to the rotating frame while holding $u_r$ and $v_r$ for the inertial frame), and it is given by 
\begin{equation}\label{eq:nk:compare:target1}
  n_{r,\, k}(\tau) 
  = \frac{1}{2} - \frac{\tilde{k}}{2\tilde{\omega}} Re \left ( \tilde{u}^*_r \tilde{v}_r \right ) - \frac{m}{4 \tilde{\omega}}\left ( |\tilde{u}_r|^2 - |\tilde{v}_r|^2 \right )~,
\end{equation}
where $\tilde{k}=  k + r \dot\phi/f$ and $\tilde{\omega} = |\tilde{\bf q}|$. One notes that the expression in Eq.~(\ref{eq:nk:compare:target1}) reproduces the result in~\cite{Adshead:2018oaa} (see Appendix B of~\cite{Adshead:2018oaa}) in the basis with the derivative coupling.  Using the relation, that connects the solutions in two frames via the field redefinition in Eq.~(\ref{eq:rot:psitoY}),
\begin{equation}
\begin{split}
  u_r &= \cos\frac{\phi}{f}\ \tilde{u}_r + i r \sin\frac{\phi}{f}\ \tilde{v}_r~,
  \\[2pt]
  v_r &= i r \sin\frac{\phi}{f}\ \tilde{u}_r + \cos\frac{\phi}{f}\ \tilde{v}_r~,
\end{split}
\end{equation}
we can also express the particle number density in the inertial frame in Eq.~(\ref{eq:nk:urvr}) in terms of $\tilde{u}_r$ and $\tilde{v}_r$, 
\begin{equation}\label{eq:nk:compare:target2}
 n_{r,\, k}(\tau) 
 = \frac{1}{2} - \frac{k}{2\omega} Re \left ( \tilde{u}_r^* \tilde{v}_r \right ) - \frac{m}{4\omega}\left ( |\tilde{u}_r|^2 - |\tilde{v}_r|^2 \right )~,
\end{equation}
where $\omega = |{\bf q}|$. It is clear from Eqs.~(\ref{eq:nk:compare:target1}) and (\ref{eq:nk:compare:target2}) that particle number densities in two frames become similar only when either $\tilde{k} \sim k$ (and thus $\tilde{\omega} \sim \omega$) or $m$ is negligible and $\tilde{k}$ is positive definite, namely $m/\omega, m/\tilde{\omega} \ll 1$ (and thus $\tilde{k}/\tilde{\omega} \sim k/\omega \sim 1$ up to the sign)~\footnote{While the approximation $m/\omega \ll 1$ for a negligible $m$ is robust, the validity of $m/\tilde{\omega} \ll 1$ depends on the situation as $\tilde{k}$ is a time-varying quantity. Similarly while $k/\omega = 1$ in the massless limit, leading to the vanishing fermion number as we expected in the inertial frame, $\tilde{k}/\tilde{\omega} = \text{sign}(\tilde{k})$ in the rotating frame (or it can flip the sign when $\tilde{k}$ becomes negative), and it can lead to the unphysical result.}. 
Apparent resemblance of the results between two frames in the middle and rightest panels of Fig.~\ref{fig:both:zeta:static} is a numerical coincidence due to a negligible $m$.

In the inflationary era, $\phi$ can satisfy the slow roll condition and its velocity is approximately constant with respect to the cosmic time $t$, or $\partial_t\phi \sim$ constant.  In this situation, one can convert the equation of motion of $\tilde{u}_r$ and $\tilde{v}_r$ in the rotating frame into the forms of Whittaker equations, and bring the solutions into those in the inertial frame via the field redefinition. When an initial boundary condition, corresponding to zero particle number, is imposed asymptotically in the far past, or $\tau \rightarrow -\infty$, it turns out that the solutions in both frames (connected via the field redefinition) simultaneously satisfy the zero particle initial condition despite their different definitions of the particle number density~\footnote{It is straightforward to understand this property by comparing Eqs.~(\ref{eq:nk:compare:target1}) and (\ref{eq:nk:compare:target2}) in the far past. In the limit of $\tau\rightarrow -\infty$, $\tilde{k} = k+ r\dot\phi/f \rightarrow k$, $\tilde{\omega} = \sqrt{(k+ r\dot\phi/f)^2 + m^2a^2} \rightarrow k$, and $\omega = \sqrt{k^2 + m^2a^2} \rightarrow k$ since $\dot\phi = a\, \partial_t \phi$ and $ma$ terms vanish. Therefore, particle number densities in two frames become identical in the far past, and one can impose the zero particle boundary condition simulataneously.}. The solutions in this case are expressed in terms of only the Whittaker function of the second kind. Instead, if one imposes the zero particle initial boundary condition at a finite initial time, the solutions include the Whittaker function of the first kind as well, and simultaneously satisfying the zero particle boundary condition in both frames is not a generic feature any more. The two unknown coefficients of the solution must be determined to satisfy the zero particle initial boundary condition in the inertial frame where the particle number is unambiguously defined.

\begin{figure}[tph]
\begin{center}
\includegraphics[width=0.47\textwidth]{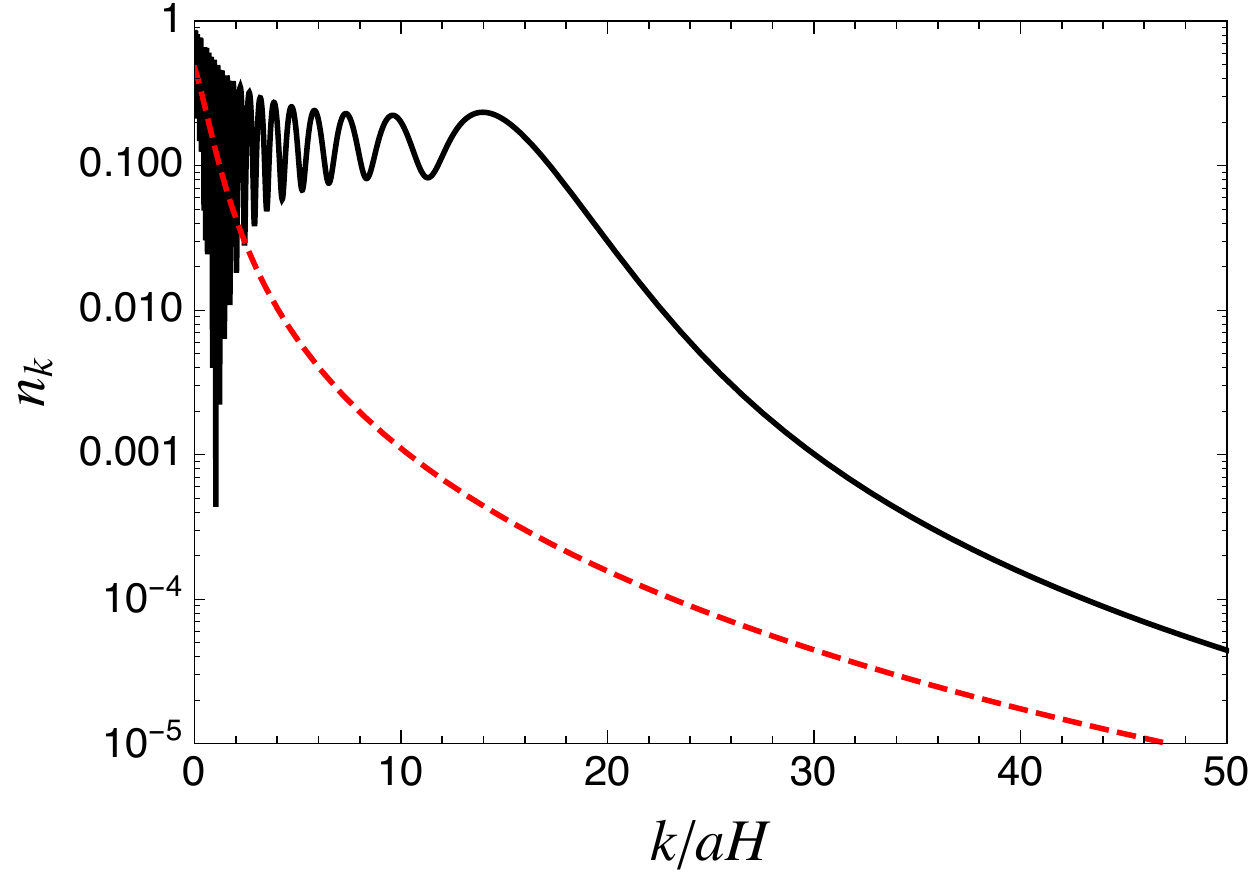}\quad
\includegraphics[width=0.47\textwidth]{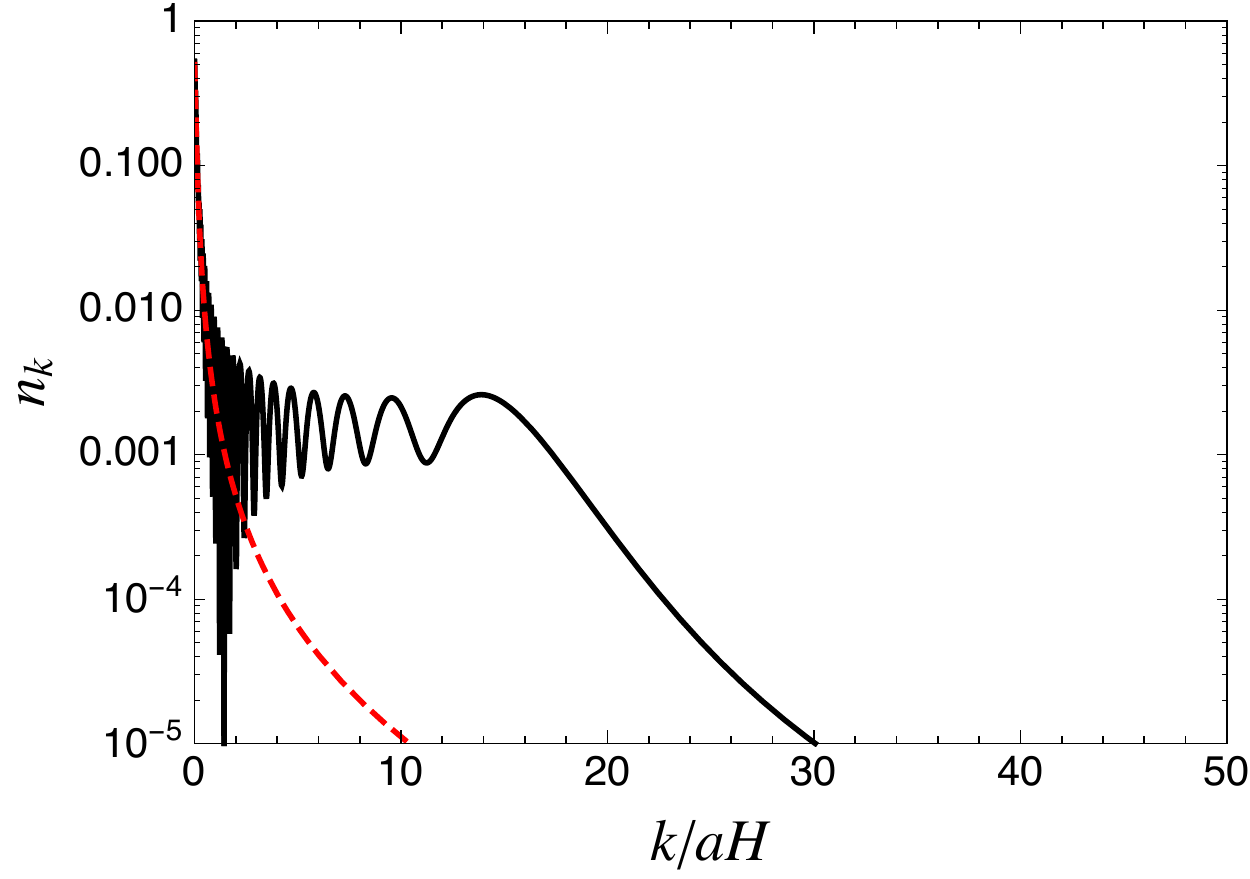}
\caption{\small The particle number density in the inertial frame for the helicity $r=-1$ (black solid) and $r=+1$ (red dashed) as a function of $k/aH$ for $\xi=10$ and $\mu=1$ (left) and $\mu = 0.1$ (right).}
\label{fig:both:zeta:inflation}
\end{center}
\end{figure}

The numerical simulation of the fermion production in the inflationary era is as straightforward as the case for the static Universe. Since $\partial_t \phi \sim$ constant (with respect to the cosmic time), the spatially homogeneous $\phi$ can be parametrized as
\begin{equation}
  \phi(\tau) = - \frac{\partial_t\phi}{H} \log \left ( \tau/\tau_{in} \right )~,
\end{equation} 
where $\tau = - 1/Ha$ ($H$ as the Hubble parameter) in de Sitter spacetime and $\tau_{in}$ is related to the initial $\phi$ value. We introduce the following set of parameters as in~\cite{Adshead:2018oaa} to elaborate our approach in a direct comparison with the literature,
\begin{equation}
  x = - k\tau~, \quad \mu = \frac{m}{H}~, \quad \xi = \frac{\partial_t \phi}{2 fH}~,
\end{equation}
and we re-express the equation for $\vec{\zeta}_r$ in the inertial frame in terms of them:
\begin{equation}\label{eq:zeta:inflation}
  \frac{1}{2} \partial_x \vec{\zeta}_r = {\bf q}_x \times \vec{\zeta}_r~,
\end{equation}
where
\begin{equation}
  {\bf q}_x = -\frac{1}{k} {\bf q} = - r\, \hat{x}_1 - \frac{\mu}{x} \sin \left ( - 4\xi \log \left ( x/x_{in} \right )  \right )\, \hat{x}_2 - \frac{\mu}{x} \cos \left ( - 4\xi \log \left ( x/x_{in} \right ) \right )\, \hat{x}_3 ~,
\end{equation}
and the vector ${\bf q}$ is given by Eq.~(\ref{eq:qvec:psi}).
All that we need to do is to evaluate the equation in Eq.~(\ref{eq:zeta:inflation}) with the initial boundary condition for $\vec{\zeta}_r$ (see Eq.~(\ref{eq:zeta:bc})), that corresponds to the zero particle number density in the far past, or $n_{r,\, k} (\tau = -\infty) = 0$,
\begin{equation}\label{eq:zeta:bc:inflation}
  \vec{\zeta}_r (\tau=-\infty) = \vec{\zeta}_r(x=\infty) = r\, \hat{x}_1~.
\end{equation}
In the numerical evaluation, $x_{in}$ (or $\tau_{in}$) can be set to any value as the fermion production does not depend on it (we have verified it through our numerical simulation). This property can be clearly understood in the rotating frame where $\phi$ derivatively couples to the fermion. Our numerical result of the particle number density in the inertial frame for the same set of parameters as those in~\cite{Adshead:2018oaa} is illustrated in Fig.~\ref{fig:both:zeta:inflation}.
As is evident in Fig.~\ref{fig:both:zeta:inflation}, our numerical simulation exactly reproduce the result in~\cite{Adshead:2018oaa} (see Fig.1 of~\cite{Adshead:2018oaa}) in a much simpler way.

\section{Summary}
\label{sec:summary}
In this work, we revisited the fermion production sourced by the classical pseudo-scalar field such as axion through its derivative coupling to the fermions. We have shown that the related dynamics can be formulated in a simpler way than the traditional approach utilizing the reparametrization group that corresponds to the freedom in selecting a representation of the gamma matrices in the Clifford algebra. 

We have established the $SU(2) \times U(1)$ subgroup (of the reparametrization group) that leaves the Clifford algebra and the Lagrangian for the Dirac fermion invariant, and that plays an essential role in our group theoretic approach. We identified the two-component column vector in a Fourier mode of the Dirac spinor which transforms like the fundamental representation of $SU(2)$ with a charge under $U(1)$. We have constructed the irreducible representations of $SO(3)\sim SU(2)$ out of the fundamental representation of $SU(2)$, and we have shown that the vectorial representation of $SO(3)$, what we called $\vec{\zeta}$ in this work, is the only non-trivial representation that one can use in our group theoretic approach.

The equation of motion in terms of $\vec{\zeta}$ turns out to be analogous to the one for a vector precessing with an angular velocity. Due to the analogy, one would expect a dictionary between the quantum-mechanical fermion production and the classical dynamics of the vector precessing with an angular velocity. The equation of motion of $\vec{\zeta}$ is universal (see Eq.~(\ref{eq:zeta:eom:generic})) irrespective of the basis choice or type of the interaction. All the details of the fermion production dynamics are entirely encoded in a quantity which corresponds to the angular velocity in the analogy. 

The particle number density was uniquely derived by a few properties in our group theoretic approach. We also explicitly derived the particle number density directly from the Hamiltonian in a simpler way in Appendix~\ref{sec:app:particlenumber}. We have argued that the particle number density (at their creation) in either inertial or non-inertial frame is at most linear in $\vec{\zeta}$ (see Eqs.~(\ref{eq:nk:zeta:generic}) and~(\ref{eq:nk:zeta:rotating})) and their apparent discrepancy is due to the different nature of the fermions in two frames in terms of their interaction. 
In the analogy to the classical dynamics, an obscure subtlety related to the transformation between two bases with and without derivative coupling of the pesudo-scalar to the fermions is translated to the physics problem that arises under the transformation from an inertial frame to the non-inertial frame, where one expects various fictitious effects, to describe the physics in a frame-independent way.

We have demonstrated through our numerical study how straightforwardly the fermion production in any situation, either static Universe or inflationary era, can be simulated in our new approach. For illustration, we have reproduced some result in literature using our new approach.

A drawback of our group theoretic formalism is that we have not managed to obtain a useful (semi) analytic expression out of the general solution for $\vec{\zeta}$ (see Eq.~(\ref{eq:zeta:exact:Tordering:generic})). A progress needs to be made on this direction. Nevertheless, there is no limitation on a numerical computation. Finally, an application of our group theoretic approach can be the extension to the production of higher-spin fermions~\cite{BasteroGil:2000je} or fermion production in an extra-dimensional spacetime.

%

\section*{Acknowledgments}
MS thanks Mohamed M. Anber, Pyungwon Ko, Zhen Liu, Adam Martin, Alex Pomarol, Alfredo Urbano for useful discussions. MS specially thanks Sergey Sibiryakov for an insightful discussion and some useful suggestions for the future work.
HS thanks June-Young Lee and Sunghyuk Park for discussions.
MS and MU were supported by the Samsung Science and Technology Foundation under Project Number SSTF-BA1602-04.

\appendix

\section{Convention}
\label{sec:app:convention}
We adopted the convention in~\cite{Adshead:2018oaa} for our explicit computations. The metric is chosen to have mostly negative signs, $\eta^{\mu\nu} = {\rm diag.}(+1,\, -1,\, -1,\, -1)$. The gamma matrices are chosen to be
\begin{equation}
  \gamma^0 = \begin{pmatrix} 1 & 0 \\[2pt] 0 & -1 \end{pmatrix}~, \quad 
  \gamma^i  = \begin{pmatrix} 0 & \sigma_i \\[2pt] -\sigma_i & 0 \end{pmatrix}~, \quad
  \gamma^5 = \begin{pmatrix} 0 & 1 \\[2pt] 1 & 0 \end{pmatrix}~.
\end{equation}

\section{Energy and Particle Number in Inertial Frame}
\label{sec:app:particlenumber}

In this section, we compute the particle number density explicitly from the Hamiltonian for the fermions in the inertial frame,
\begin{equation}
\begin{split}
 \mathcal{H} 
  &= \bar{\psi} \left ( - i \gamma^i \partial_i + m_R - i\, m_I\, \gamma^5 \right ) \psi~,
\end{split}
\end{equation}
and we confirm that the result agrees with what we obtained in Eq.~(\ref{eq:nk:zeta:generic}).

We express the Hamiltonian operator in the tensor product form,
\begin{equation}
 h_D \equiv -i \gamma^i \partial_i + m_R - i\, m_I \gamma^5 = i\, \sigma_2 \otimes (\vec{\sigma} \cdot {\bf k}) + m_R I_{\bf 2} \otimes I_{\bf 2} - i\, m_I \sigma_1 \otimes I_{\bf 2}~.
\end{equation}
We also express the fermion quantum field in the form such that the action of the $h_D$ operator is manifest. The fermion quantum field is given by
\begin{equation}\label{eq:app:quantum:psi}
\begin{split}
  \psi = \int \frac{d^3k}{(2\pi)^{3/2}} e^{i{\bf{k}}\cdot {\bf{x}}} \sum_{r=\pm} \left [ U_r( {\bf{k}}, \tau) a_r ({\bf{k}}) + V_r(-{\bf{k}},\tau) b^\dagger_r (-{\bf{k}})  \right ]~,
\end{split}
\end{equation}
where
\begin{equation}
  U_r({\bf k}, \tau) = \frac{1}{\sqrt{2}} \begin{pmatrix} u_r\, \chi_r \\ r v_r \chi_r   \end{pmatrix}~, \quad
  V_r ({\bf k}, \tau) = C\bar{U}^T_r \quad \hbox{with}\quad C = \begin{pmatrix}  0 & i \sigma_2 \\ i \sigma_2 & 0 \end{pmatrix}
  = i\sigma_1 \otimes \sigma_2~,
\end{equation}
and 
\begin{equation}
 \chi_r ({\bf k}) = \frac{\left ( k + r \vec\sigma \cdot {\bf k} \right )}{\sqrt{2k (k+ k_3)}} \bar{\chi}_r~, \quad
 \bar{\chi}_+ = \begin{pmatrix} 1 \\ 0 \end{pmatrix}~,\quad
 \bar{\chi}_- = \begin{pmatrix} 0 \\ 1 \end{pmatrix}~.
\end{equation}
The part inside $[\ \ \ \ ]$ in the fermion quantum field $\psi$ in Eq.~(\ref{eq:app:quantum:psi}) can be written in a tensor product form,
\begin{equation}\label{eq:app:UrVr:psi}
\begin{split}
  & U_r( {\bf{k}}, \tau) a_r ({\bf{k}}) + V_r(-{\bf{k}},\tau) b^\dagger_r (-{\bf{k}}) \\[2pt]
  & = \left ( \xi_r({\bf k}) \otimes \chi_r ({\bf k})\right ) a_r ({\bf k}) + \left ( i\, r\, \sigma_2\, \xi_r^* (-{\bf k}) \otimes \chi_{-r} (-{\bf k})\right ) b^\dagger_r (-{\bf k})~,\\[2pt]
  & = \left ( \xi_r({\bf k}) \otimes \chi_r ({\bf k})\right ) a_r ({\bf k}) + \left ( e^{-i\, r \varphi_{\bf k}}\, i\, \sigma_2\, \xi_r^* (-{\bf k}) \otimes \chi_{r} ({\bf k})\right ) b^\dagger_r (-{\bf k})~.\\
\end{split}
\end{equation}
In the last line of Eq.~(\ref{eq:app:UrVr:psi}), we have expressed two-component spinor in terms on $\chi_r({\bf k})$ using the relation,
\begin{equation}
\begin{split}
\chi_{-r} (-{\bf k}) = r e^{-i\, r\varphi_{\bf k}} \chi_r ({\bf k})\quad \hbox{with} \quad e^{i\varphi_{\bf k}} = \frac{k_1 + i\, k_2}{\sqrt{k_1^2 + k_2^2}}~,
\end{split}
\end{equation}
and this will help simplifying the computation. We also hid the time dependence in Eq.~(\ref{eq:app:UrVr:psi}) (and in what follows) for a notational simplicity. We act the Hamiltonian operator on the individual terms in front of $a_r$ and $b_r^\dagger$ in Eq.~(\ref{eq:app:UrVr:psi}). Using the relations, $(\vec{\sigma} \cdot {\bf k}) \chi_r ({\bf k}) = rk\, \chi_r ({\bf k})$, we obtain
\begin{equation}\label{eq:app:hDUr:hDVr}
\begin{split}
h_D \left ( \xi_r({\bf k}) \otimes \chi_r ({\bf k}) \right )
  &= \Big [ \left ( i\, rk\, \sigma_2  + m_R I_{\bf 2}  - i\, m_I \sigma_1 \right ) \otimes I_{\bf 2} \Big ] \left ( \xi_r({\bf k}) \otimes \chi_r ({\bf k}) \right )~,\\[3pt]
  &=\Big [ \sigma_3 \left ( {\bf q}\cdot \vec{\sigma} \right ) \otimes I_{\bf 2} \Big ] \left ( \xi_r({\bf k}) \otimes \chi_r ({\bf k}) \right )~,\\[5pt]
h_D \left (i\, r\, \sigma_2\, \xi^*_r(-{\bf k}) \otimes \chi_{-r} (-{\bf k}) \right )
  &= e^{-i\, r \varphi_{\bf k}}  \Big [  \left ( -rk I_{\bf 2}  +i\, m_R \sigma_2  + i\, m_I \sigma_3 \right ) \otimes I_{\bf 2} \Big ] \left ( \xi^*_r(-{\bf k}) \otimes \chi_{r} ({\bf k}) \right ) ~,\\[3pt]
  &= e^{-i\, r \varphi_{\bf k}} \Big [ \sigma_1 \left ( {\bf q}_0\cdot \vec{\sigma} \right ) \otimes I_{\bf 2} \Big ] \left ( \xi^*_r(-{\bf k}) \otimes  \chi_{r} ({\bf k}) \right )~ , 
\end{split}
\end{equation}
where ${\bf q}_0 = - rk\, \hat{x}_1 + m_I\, \hat{x}_2 - m_R\, \hat{x}_3$ was introduced purely for the simple expression.

What we computed in Eq.~(\ref{eq:app:hDUr:hDVr}) will be multiplied by the corresponding terms in $\bar{\psi}$ from the left.
The $\bar\psi = \psi^\dagger \gamma^0$ (with $\gamma^0 = \sigma_3 \otimes I_{\bf 2}$) includes the pieces,
\begin{equation}
\begin{split}
&\big [ a^\dagger_r ({\bf k})\, U^\dagger_r ({\bf k}, t) + b_r (-{\bf{k}})\, V^\dagger_r(-{\bf{k}},t) \big ] \left ( \sigma_3 \otimes I_{\bf 2} \right ) \\[2pt]
=&~\big [ a^\dagger_r ({\bf k})\left ( \xi^\dagger_r ({\bf k}) \otimes \chi^\dagger_r ({\bf k}) \right )+ b_r (-{\bf{k}})\left ( e^{i\, r\varphi_{\bf k}}\,  \xi^T_r(-{\bf{k}}) (-i\, \sigma_2) \otimes \chi^\dagger_r ({\bf k}) \right ) \big ] \left ( \sigma_3 \otimes I_{\bf 2} \right ) \\[2pt]
=&~ a^\dagger_r ({\bf k})\left ( \xi^\dagger_r ({\bf k})\, \sigma_3 \otimes \chi^\dagger_r ({\bf k}) \right )+ b_r (-{\bf{k}})\left ( e^{i\, r\varphi_{\bf k}}\,  \xi^T_r(-{\bf{k}})\, \sigma_1 \otimes \chi^\dagger_r ({\bf k}) \right )~.
\end{split}
\end{equation}
In the computation of the Hamiltonian in terms of the creation and annihilation operators, we can forget about two-component spinor parts from now on. It will drop due to the orthonormality of $\chi_r({\bf k})$.
The resulting Hamiltonian in terms of the creation and annihilation operators can be written as
\begin{equation}\label{eq:app:Eenergy:operators}
  {\mathcal{H}} = \sum_{r=\pm}\int d^3k \left ( a^\dagger_r ({\bf k}),\, b_r (-{\bf k})  \right ) \begin{pmatrix} A_r & B^*_r \\[2pt] B_r & - A_r\end{pmatrix} \begin{pmatrix} a_r ({\bf k}) \\[2pt]  b^\dagger_r (-{\bf k})  \end{pmatrix}~,
\end{equation}
where the matrix elements are easily computed using all the ingredients that we prepared before Eq.~(\ref{eq:app:Eenergy:operators}), and they are given by
\begin{equation}\label{eq:app:Ar:Br}
\begin{split}
  A_r =&~  \xi_r^\dagger ({\bf k})\left ( {\bf q}\cdot \vec{\sigma} \right ) \xi_r ({\bf k}) = {\bf q}\cdot \vec{\zeta}_r~,\\[2pt]
  =&~ \frac{1}{2} \Big [ k \left ( u^*_r v_r + v^*_r u_r \right ) - i\, r\, m_I \left ( u^*_r v_r - v^*_r u_r \right ) + m_R \left ( |u_r|^2 - |v_r|^2 \right )   \Big ]~,\\[3pt]
  B_r =&~ \left ( e^{i\, r\varphi_{\bf k}}\,  \xi^T_r(-{\bf{k}})\, \sigma_1  \right ) \sigma_3 \left ( {\bf q}\cdot \vec{\sigma} \right ) \xi_r ({\bf k}) 
  =  -i\, e^{i\, r\varphi_{\bf k}}\,  \xi^T_r(-{\bf{k}})\, \sigma_2 \left ( {\bf q}\cdot \vec{\sigma} \right ) \xi_r ({\bf k})~,\\[2pt]
  =&~ \frac{r e^{i\, r\varphi_{\bf k}}}{2} \Big [ 2 m_R u_r v_r - k \left ( u^2_r - v^2_r \right ) - i\, r m_I \left ( u^2_r + v^2_r \right ) \Big ]~.
\end{split}
\end{equation}
We wrote the expressions of $A_r$ and $B_r$ in terms of $u_r$ and $v_r$ (and their complex conjugates) in Eq.~(\ref{eq:app:Ar:Br}) only for the purpose of comparison with literature, and they agree with those in~\cite{Adshead:2018oaa}. Due to the non-vanishing matrix element $B_r$, the $a^\dagger_r$ and $b^\dagger_r$ (and  $a_r$ and $b_r$) do not create (and destroy) the energy eigenstates. Diagonalizing the matrix in Eq.~(\ref{eq:app:Eenergy:operators}) gives rise to two eigenvalues, $\pm \omega$, with $\omega = |{\bf q}| = \sqrt{k^2 + m^2_R + m^2_I}$, or
\begin{equation}\label{eq:app:matrix:diagonalize}
  \begin{pmatrix} A_r & B^*_r \\[2pt] B_r & - A_r\end{pmatrix} = 
  \begin{pmatrix} \alpha^*_r & \beta^*_r \\[2pt] -\beta_r & \alpha_r\end{pmatrix} 
  \begin{pmatrix} \omega & 0  \\[2pt] 0 & - \omega \end{pmatrix}
  \begin{pmatrix} \alpha_r & -\beta^*_r \\[2pt] \beta_r & \alpha^*_r\end{pmatrix}~.
\end{equation}
Since the matrix elements, $A_r$ and $B_r$ are quadratic in $u_r$ and $v_r$, the mixing angles, $\alpha_r$ and $\beta_r$, must be linear in $u_r$ and $v_r$.
Expressing the Hamiltonian in Eq.~(\ref{eq:app:Eenergy:operators}) in terms of the creation and annihilation operators amounts to
\begin{equation}\label{eq:app:rotation}
 \begin{pmatrix} a_r ({\bf k}) \\[2pt]  b^\dagger_r (-{\bf k})  \end{pmatrix} 
 \rightarrow 
 \begin{pmatrix} \alpha^*_r & \beta^*_r \\[2pt] -\beta_r & \alpha_r\end{pmatrix}
 \begin{pmatrix} a_r ({\bf k}) \\[2pt]  b^\dagger_r (-{\bf k})  \end{pmatrix}~.
\end{equation}
Therefore, the particle number density (with the diagonalized $a_r$ and $a_r^\dagger$) for a particle with the helicity $r$ is given by
\begin{equation}
  N_r(\tau) \equiv \langle 0| \int \frac{d^3 k}{(2\pi)^3}\, a_r^\dagger\, a_r | 0\rangle  = \int d^3 k\, n_{r,\, k} (\tau)  = \int d^3 k\, |\beta_r|^2~.
\end{equation}
The particle number density for a anti-particle is also given by $|\beta_r|^2$.

Since $\beta_r$ is linear in $u_r$ and $v_r$, it does not have a simple expression in terms of $\vec{\zeta}_r$. However, what we need to know is $|\beta_r|^2$ which could be, in principle, a linear function in $\vec{\zeta}_r$. Indeed, from the Eq.~(\ref{eq:app:matrix:diagonalize}), 
\begin{equation}
 A_r = \omega \left ( |\alpha_r|^2 - |\beta_r|^2\right ) = \omega \left ( 1 - 2 |\beta_r|^2\right )\quad \rightarrow \quad
  |\beta_r|^2 = \frac{1}{2} \left ( 1 - \frac{A_r}{\omega} \right )~,
\end{equation}
where we used $|\alpha_r|^2+|\beta_r|^2 = 1$. Using the $A_r$ in terms of $\vec{\zeta}_r$ in Eq.~(\ref{eq:app:Ar:Br}) and the relation, $\omega = |{\bf q}|$, we finally derive the particle number density to be
\begin{equation}\label{eq:app:nk:final}
 n_{r,\, k}(\tau)  = \frac{1}{2} \left ( 1 - \frac{{\bf q}\cdot \vec{\zeta}_r}{|{\bf q}|} \right )~. 
\end{equation}
The ${\bf q}\cdot \vec{\zeta}_r$ is the diagonal element of the Hamiltonian matrix in Eq.~(\ref{eq:app:Eenergy:operators}), and it can be thought of as the energy eigenvalue when the Hamiltonian matrix is diagonal, or $B_r = 0$ in Eq.~(\ref{eq:app:nk:final}). After a simple algebra (eigenvalue equation), one can derive the relation,
\begin{equation}\label{eq:app:vanishingBr}
  |B_r|^2 = \omega^2 - \big ( {\bf q}\cdot \vec{\zeta}_r \big )^2 = {\bf q}\cdot {\bf q} - \big ( {\bf q}\cdot \vec{\zeta}_r \big ) \, \big ( {\bf q}\cdot \vec{\zeta}_r \big ) = \big ( {\bf q} \times \vec{\zeta}_r \big )^2~,
\end{equation}
which shows that the off-diagonal element of the Hamiltonian vanishes, $B_r =0$, when $\vec{\zeta}_r$ is parallel or anti-parallel to the vector ${\bf q}$. We restrict the discussion below to the case where $\vec{\zeta}_r$ is parallel or anti-parallel to ${\bf q}$ for a clarity.

What we compute for the particle number density sourced by the classical $\phi$-field is the vacuum expectation value, $\langle 0 | a^\dagger_r a_r |0\rangle$, instead of the expectation value of the number operator with a generic state vector (similarly for the energy expectation values). When $\vec{\zeta}_r$ is parallel to the ${\bf q}$ vector, ${\bf q}\cdot \vec{\zeta}_r$ becomes an energy eigenvalue of the Hamiltonian, while $\langle 0 | a^\dagger_r a_r |0\rangle$ vanishes since $a_r$ becomes an annihilation operator defining the vacuum, which implies that there is no energy released into the vacuum to create particles. As a time goes on, the Hamiltonian becomes non-diagonal, and the vacuum $|0\rangle$ in the definition of the particle number (see Eq.~(\ref{eq:N:define})) does not corresponds to the truth vacuum from which one-particle states are generated. This is how a nonzero fermion production can happen. When $\vec{\zeta}_r$ reaches the configuration which is anti-parallel to the ${\bf q}$ vector, ${\bf q}\cdot \vec{\zeta}_r$  becomes an energy eigenvalue of the Hamiltonian again (nevertheless, the energy of the Hamiltonian stay same as before due to the exchanged roles between $a_r$ ($b_r^\dagger$) and $b_r^\dagger$ ($a_r$)), and the particle number density becomes maximum, $\langle 0 | a^\dagger_r a_r |0\rangle = |\beta_r|^2 \langle 0 | b_r b_r^\dagger |0\rangle = |\beta_r|^2$, via the maximal mixing, $a_r = \beta_r^*\, b_r^\dagger$, in Eq.~(\ref{eq:app:rotation}). 





\bibliography{lit}

\end{document}